\documentclass[useAMS,usenatbib,usegraphicx]{mn2e}
\usepackage{graphicx}

\newcommand{\be}{\begin{equation}}
\newcommand{\ee}{\end{equation}}

\newcommand{\bl}[1]{\mbox{\boldmath$ #1 $}}

\newcommand{\aleff}{\alpha_{\rm eff}} 

\begin{document}

\title[Secular evolution of circumstellar discs]{Secular evolution of viscous and self-gravitating circumstellar discs}
\author[E. I. Vorobyov and S. Basu]{E. I. Vorobyov$^{1,2}$\thanks{E-mail:
vorobyov@ap.smu.ca (EIV); basu@astro.uwo.ca (SB)}, Shantanu Basu$^{3}$ \\
$^{1}$The Institute for Computational Astrophysics, Saint Mary's University, 
Halifax, NS, B3H 3C3, Canada \\
$^{2}$Institute of Physics, South Federal University, Rostov-on-Don, 344090, Russia \\
$^{3}$Department of Physics and Astronomy, University of Western Ontario, London, Ontario,
N6A 3K7, Canada}

\maketitle

\begin{abstract}
We add the effect of turbulent viscosity via the $\alpha-$prescription
to models of the self-consistent formation and evolution of protostellar
discs. 
Our models are non-axisymmetric and carried out using the thin-disc 
approximation. 
Self-gravity plays an important role in the early evolution of a disc, 
and the later evolution is determined by the relative importance of
gravitational and viscous torques.
In the absence of viscous torques, a protostellar disc evolves
into a self-regulated state with disk-averaged Toomre parameter $Q \sim 1.5-2.0$, 
non-axisymmetric structure diminishing with time, and maximum disc-to-star mass ratio
$\xi = 0.14$. We estimate an effective viscosity parameter 
$\alpha_{\rm eff}$ associated with gravitational torques 
at the inner boundary of our simulation to be in the range 
$10^{-4}-10^{-3}$ during the late evolution.
Addition of viscous torques with a low value $\alpha = 10^{-4}$ 
has little effect on the evolution, structure, and accretion 
properties of the disc, and the self-regulated
state is largely preserved.
A sequence of increasing values of $\alpha$ results in the discs becoming
more axisymmetric in structure, being more gravitationally stable, having
greater accretion rates, larger sizes, shorter lifetimes, and 
lower disc-to-star mass ratios. 
For $\alpha=10^{-2}$, the model is viscous-dominated and 
the self-regulated state largely disappears by late times. The axisymmetry and low
surface density of this model
may contrast with observations and pose problems for planet formation models.
The use of $\alpha=0.1$ leads to very high disc accretion 
rates and rapid (within 2 Myr) depletion of the disc, and 
seems even less viable observationally.
Furthermore, only the non-viscous-dominated
models with low values of $\alpha = 10^{-4}-10^{-3}$
can account for an early phase of quiescent low accretion rate
$\dot{M} \sim 10^{-8}~M_\odot$~yr$^{-1}$ (interspersed with accretion bursts)
that can explain the recently observed Very Low luminosity Objects (VeLLOs).
We also find that a modest increase in disc temperature caused by a stiffer
barotropic equation of state ($\gamma=1.67$) has little effect on
the disc accretion properties averaged over many disc orbital periods ($\sim 10^4$~yr),
but can substantially influence the instantaneous mass accretion rates, 
particularly in the early embedded phase of disc evolution.
\end{abstract}
\begin{keywords}
accretion, accretion discs û- hydrodynamics û- instabilities û- ISM: clouds û- stars:
formation
\end{keywords} 

\section{Introduction}
We have recently demonstrated that disc gravity 
plays an important role not only in the early embedded phase of disc evolution but also
in the late accretion phase \citep{VB3,VB4}. In the early embedded phase, when the
infall of matter from the surrounding envelope is substantial, 
mass is transported inward by the gravitational torques from spiral arms that are
a manifestation of the envelope-induced gravitational instability in
the disc. In the late accretion phase, when the
gas reservoir of the envelope is depleted, the distinct spiral structure 
is replaced by ongoing irregular nonaxisymmetric density perturbations. 
These perturbations produce a residual nonzero gravitational torque in the disc.
In particular, the net gravitational torque in the inner disc tends
to be negative during first several million years of the evolution, while
the outer disc has a net positive gravitational torque.  This is a fundamental 
property of self-gravitating circumstellar discs around low-mass stars. 
There is also an overall net negative torque in the disc that is related
to the removal of angular momentum by gas that is accreted in to the
central object.
Although we do not model the gas flow within a central sink of size
5 AU, the angular momentum that is carried in to this region 
may be lost to the system via additional processes including magnetic braking and outflows,
thereby allowing the gas to reach the central star. 

In this paper we seek to determine the effect of {\it other} mechanisms of radial mass 
and angular momentum transport on the secular evolution of self-consistently formed 
circumstellar discs. It is well known that standard collisional viscosity (molecular viscosity) 
is negligible in application to circumstellar discs.
The best candidate to date is turbulent viscosity induced by the magneto-rotational instability
(MRI) \citep{BH}, though other mechanisms such as nonlinear hydrodynamic turbulence cannot 
be completely eliminated due to the large Reynolds numbers involved \citep[e.g.][]{Afshordi}. 
We make no specific assumptions about the source of turbulence and 
parameterize the magnitude of turbulent viscosity using the usual 
$\alpha$-prescription \citep{SS}
\begin{equation}
\nu=\alpha \,c_{s}\,H,
\label{alpha}
\end{equation} 
where $c_{s}$ is the sound speed and $H$ is the disc scale height. 
The most appropriate value of the parameter $\alpha$ in circumstellar discs is uncertain.
Both the shearing box and global numerical simulations
of the MRI tend to yield the values of $\alpha$ that vary significantly from model to model
and range between $10^{-4}$ and ${\rm a~few} \times 10^{-1}$ \citep[e.g.][]{Hawley95,
Fleming,Brandenburg96,Stone96,Armitage}. Motivated by these studies, we have adopted a
similar range of values for $\alpha$. We also assume that $\alpha$ is spatially and temporally 
constant, i.e. it represents a mean value, time-averaged over many orbital periods of the disc.
Radial variations in $\alpha$ may (and should) be present in the disc but it requires a more 
thorough consideration of the disc physics and is left for a follow-up
paper. 

Our study of the self-consistent formation and long-term evolution of circumstellar discs
has been preceded by numerical simulations of \citet{Lin,Nakamoto95,Hueso05} and others.
However, our work is different in one important aspect -- we employ a fully two-dimensional 
numerical hydrodynamics simulations in the thin-disc approximation 
(in contrast to earlier one-dimensional studies). As a result, 
we account for disc self-gravity self-consistently by solving the Poisson integral 
\citep[see][]{VB2} and need not parameterize gravitational torques in terms of 
the $\alpha$-prescription. Use of the thin-disc approximation is, unfortunately,
a necessity. Fully three-dimensional numerical simulations are too computationally 
expensive to study the disc evolution on time scales of several Myr.

The paper is organized as follows. Section~\ref{model} gives a brief description of model equations
and initial conditions. The main results are presented in Section~\ref{results}. The observational
implications and comparison with earlier studies are discussed in Section~\ref{discuss}.
The main conclusions are summarized in Section~\ref{summary}.

\section{Model description}
\label{model}
We use the thin-disc approximation to compute the evolution of
non-axisymmetric rotating, gravitationally bound cloud cores. We start our numerical 
integration in the pre-stellar phase, which is characterised by a collapsing {\it starless}
cloud core, and continue into the late accretion phase, which  is characterised
by a protostar/disc system. This ensures a {\it self-consistent} formation 
of circumstellar discs in our numerical simulations. 
Once the disc is formed, its subsequent evolution is determined by an interplay 
between the efficiency of the mass and angular momentum transport in the 
disc\footnote{In fact, discs may also transport
angular momentum to the external environment due to magnetic braking. This effect will be 
considered in a follow-up paper.} and the infall rate of matter from the 
surrounding envelope onto the disc. The disc-envelope interaction is taken into account self-consistently,
since we evolve numerically the disc and envelope altogether. It means that there is no
source term in the numerical grid allowing for mass deposition from the envelope, but the mass 
infall rate onto the disc is actually determined by
the dynamics of gas in the envelope. The disc occupies the innermost regions of our numerical grid and
the envelope occupies the rest of the grid. 
We note the infall rate of matter from the envelope onto the disc is not necessarily the same 
as the mass accretion rate from the disc onto the protostar. While the former shows a fast decline
with time, the latter is usually characterized by a much slower decline and has a strong dependence
on the stellar mass \citep{VB3,VB4}.

For details of the basic equations, numerical methods and numerical tests
we refer the reader to \citet{VB2}. Here we briefly provide
the basic equations modified to include the effect of viscosity.
The basic equations of mass and momentum transport in the thin-disc approximation are
\begin{eqnarray}
\label{cont}
 \frac{{\partial \Sigma }}{{\partial t}} & = & - \nabla _p  \cdot \left( \Sigma \bl{v}_p 
\right), \\ 
\label{mom}
 \Sigma \frac{d \bl{v}_p }{d t}  & = &  - \nabla _p {\cal P}  + \Sigma \, \bl{g}_p + 
 (\nabla \cdot \mathbf{\Pi})_p \, ,
\end{eqnarray}
where $\Sigma$ is the mass surface density, ${\cal P}=\int^{Z}_{-Z} P dz$ is the vertically integrated
form of the gas pressure $P$, $Z$ is the radially and azimuthally varying vertical scale height,
$\bl{v}_p=v_r \hat{\bl r}+ v_\phi \hat{\bl \phi}$ is the velocity in the
disc plane, $\bl{g}_p=g_r \hat{\bl r} +g_\phi \hat{\bl \phi}$ is the gravitational acceleration 
in the disc plane, and $\nabla_p=\hat{\bl r} \partial / \partial r + \hat{\bl \phi} r^{-1} 
\partial / \partial \phi $ is the gradient along the planar coordinates of the disc. 
We note that $\bl{g}_p$ includes the input from the central star when it forms.
The viscous stress tensor $\mathbf{\Pi}$ is expressed as
\begin{equation}
\mathbf{\Pi}=2 \Sigma\, \nu \left( \nabla v - {1 \over 3} (\nabla \cdot v) \mathbf{e} \right),
\end{equation}
where $\nabla v$ is a symmetrized velocity gradient tensor, $\mathbf{e}$ is the unit tensor, and
$\nu$ is the kinematic viscosity. 
The components of $(\nabla \cdot \mathbf{\Pi})_p$ in polar coordinates ($r,\phi$) 
are given in the Appendix. We emphasize that we do not take any simplifying assumptions about 
the form of the viscous stress tensor, apart from those imposed by the adopted 
thin-disc approximation. 
It can be shown \citep{Lodato08} that equation~(\ref{mom}) can
be reduced to the usual equation for the conservation of angular momentum of a radial annulus 
in the axisymmetric viscous accretion disc \citep{Pringle}.

Equations~(\ref{cont}) and (\ref{mom}) are closed with a barotropic equation
that makes a smooth transition from isothermal to adiabatic evolution at 
$\Sigma = \Sigma_{\rm cr} = 36.2$~g~cm$^{-2}$:
\begin{equation}
{\cal P}=c_s^2 \Sigma +c_s^2 \Sigma_{\rm cr} \left( \Sigma \over \Sigma_{\rm cr} \right)^{\gamma},
\label{barotropic}
\end{equation}
where $c_s=0.188$~km~s$^{-1}$ is the sound speed in the beginning of numerical simulations 
and $\gamma=1.4$. Equation~(\ref{barotropic}), though neglecting detailed
cooling and heating processes, was shown to reproduce to a first approximation 
the radial temperature gradients in circumstellar discs \citep{VB3} 
and the density-temperature relation in collapsing cloud cores \citep{VB2}. 

It should be stressed here that circumstellar discs described by the 
barotropic equation of state with $\gamma=1.4$ are susceptible to fragmentation and 
formation of stable clumps in the early embedded phase of evolution.
A more accurate treatment of the energy balance in the disc 
involving radiative cooling from the disc surface and shock heating due to artificial 
viscosity has shown that the strength of gravitational instability in general and the disc propensity to fragmentation in particular depend on the rate of cooling 
\citep[e.g.][]{Gammie01,Johnson03,Rice03,Lodato04,Mejia05}. In addition, 
fragmentation can be stabilized 
in the inner discs by slow cooling \citep[e.g.][]{Stamatellos} and in the outer 
discs by stellar and envelope irradiation \citep{Matzner05,Cai08}. However, recent 
semi-analytical and numerical studies (including radiation transfer) 
reveal clump formation in discs (particularly, in their outer parts) 
around stars with mass equal to or more massive than one solar mass
\citep{Krumholz07,Mayer07,Stamatellos07,Boss08,Kratter08}. 
These simulations produce discs that are usually hotter than those described 
by a barotropic equation of state with $\gamma=1.4$
but that does not necessarily imply less efficient transport due to gravitational torques.
As \citet{Cai08} have demonstrated, a modest rise in the disc temperature (from the envelope irradiation)
may in fact promote transport due to a growing relative strength of low-order spiral modes 
($m\le 2$) in the disc.
In order to examine if a higher disc temperature can affect our main results,
we  consider a stiffer barotropic equation of state with $\gamma=1.67$ in Section~\ref{hot}.

The kinematic viscosity is computed during numerical simulations as 
\begin{equation}
\label{viscosity}
\nu = \alpha \, \tilde{c}_{s} \, Z,
\end{equation}
where $\tilde{c}^2_{s}=\partial {\cal P} /\partial \Sigma$ is the effective sound speed
of (generally) non-isothermal gas. The vertical scale height $Z$ is determined in
each computational cell using
an assumption of local hydrostatic equilibrium in the gravitational field of
the central star and the disc (see Appendix A).

The initial conditions are similar to those in \citet{VB3}.
The initial radial surface density and angular velocity
profiles of the model cloud core with mass $0.8~M_\odot$ and mean molecular weight 
$2.33$ are characteristic of a collapsing axisymmetric magnetically supercritical core \citep{Basu}:
\begin{equation}
\Sigma={r_0 \Sigma_0 \over \sqrt{r^2+r_0^2}}\:,
\label{dens}
\end{equation}
\begin{equation}
\Omega=2\Omega_0 \left( {r_0\over r}\right)^2 \left[\sqrt{1+\left({r\over r_0}\right)^2
} -1\right].
\end{equation}
The scale length $r_0 = k c_s^2 /(G\Sigma_0)$, where $k= \sqrt{2}/\pi$
and $\Sigma_0=0.12$~g~cm$^{-2}$. The central angular velocity is $\Omega_0=1.1$~km~s$^{-1}$~pc$^{-1}$.
We have adopted a somewhat higher value of $\Omega_0$ than in \citet{VB3} 
in order to emphasize the burst phase of mass accretion.
Our adopted initial profiles are characterized by the important
dimensionless free parameter $\eta \equiv  \Omega_0^2r_0^2/c_s^2$.
The asymptotic ($r \gg r_0$) ratio of centrifugal to gravitational
acceleration has magnitude $\sqrt{2}\,\eta$ \citep[see][]{Basu} and
the centrifugal radius of a
mass shell initially located at radius $r$ is estimated to be
$r_{\rm cf} = j^2/(Gm) = \sqrt{2}\, \eta r$.
For our chosen parameters, we find $\eta=1.42 \times 10^{-3}$, 
and $r_{\rm cf} =16.6$ AU for the outermost mass shell located initially
at $r_{\rm out} = 0.04$ pc.

Equations~(\ref{cont}), (\ref{mom}), (\ref{barotropic}) are solved in polar 
coordinates $(r, \phi)$ on a numerical grid with
$128 \times 128$ points. We use the method of finite differences with a time-explicit,
operator-split solution procedure. Advection is
performed using the second-order van Leer scheme.  The radial points are logarithmically spaced.
The innermost grid point is located at $r=5$~AU, and the size of the 
first adjacent cell is 0.3~AU.  We introduce a ``sink cell'' at $r<5$~AU, 
which represents the central star plus some circumstellar disc material, 
and impose a free inflow inner boundary condition. The outer boundary is reflecting.
The gravity of a thin disc is computed by directly summing the input from each computational cell
to the total gravitational potential. The convolution theorem is used to speed up 
the summation. A small amount of artificial viscosity is added to the code, 
though the associated artificial viscosity torques were shown to be negligible 
in comparison with gravitational torques \citep{VB3}. 
A more detailed explanation of numerical methods and relevant tests 
can be found in \citet{VB2,VB3}.

\section{Results}
\label{results}
We consider five models, each having identical initial conditions but distinct
values of spatially and temporally uniform $\alpha$. In particular, model~1 is characterized 
by $\alpha=0$
and is used as the standard model against which other models with non-zero viscosity are compared.
Models~2, 3, 4, and 5 have $\alpha$ equal to $10^{-4}$, $10^{-3}$, $10^{-2}$, and $10^{-1}$, 
respectively. To facilitate the comparison between viscous and non-viscous models, 
$\alpha$ is kept zero in the early evolution in all models and is
set to its corresponding value only after the disc is formed at $t\approx 0.14$~Myr.  

\begin{figure}
  \resizebox{\hsize}{!}{\includegraphics{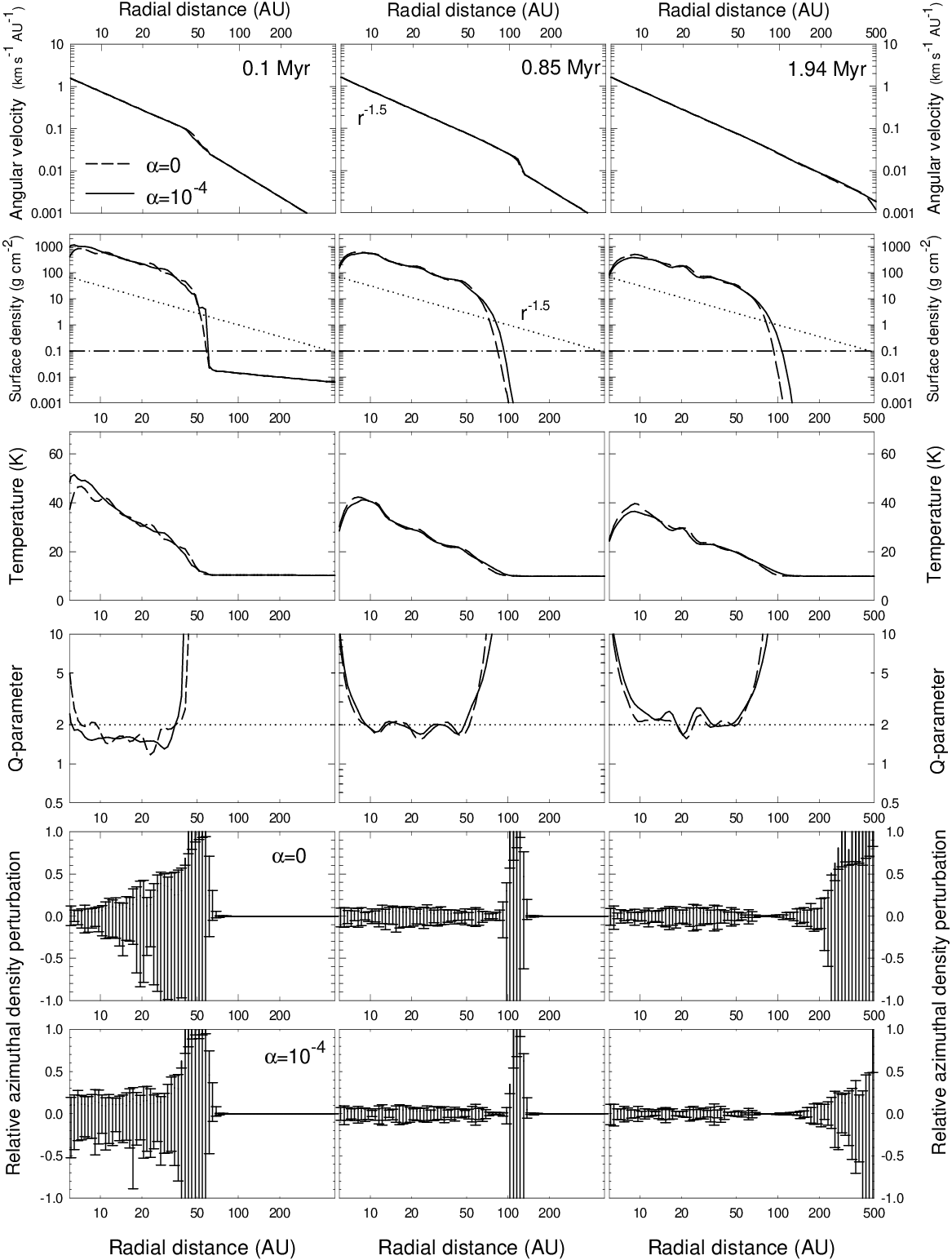}}
      \caption{Radial structure of the circumstellar disc in the $\alpha=0$ model~1 (dashed lines) 
      and $\alpha=10^{-4}$ model~2 (solid
      lines). Top to bottom: azimuthally averaged values of $\overline{\Omega}$, $\overline{\Sigma}$,
      $\overline{T}$, $\overline{Q}$, and $\Delta \Sigma$ in the $0.1$-Myr-old disc (left-hand column),
      $0.85$-Myr-old disc (middle column) and $1.94$-Myr-old disc (right-hand column). 
      The error bars in two bottom rows show the minimum and maximum $\Delta \Sigma(r_i,\phi_i)$ 
      in each radial annulus. The dotted lines
      in the second row show the surface density as inferred from the minimum mass solar nebula model, 
      and the dot-dashed lines mark a fiducial critical density ($1.0$~g~cm$^{-2}$) for transition 
      between the disc and envelope.}
         \label{fig1}
\end{figure}

\begin{figure}
  \resizebox{\hsize}{!}{\includegraphics{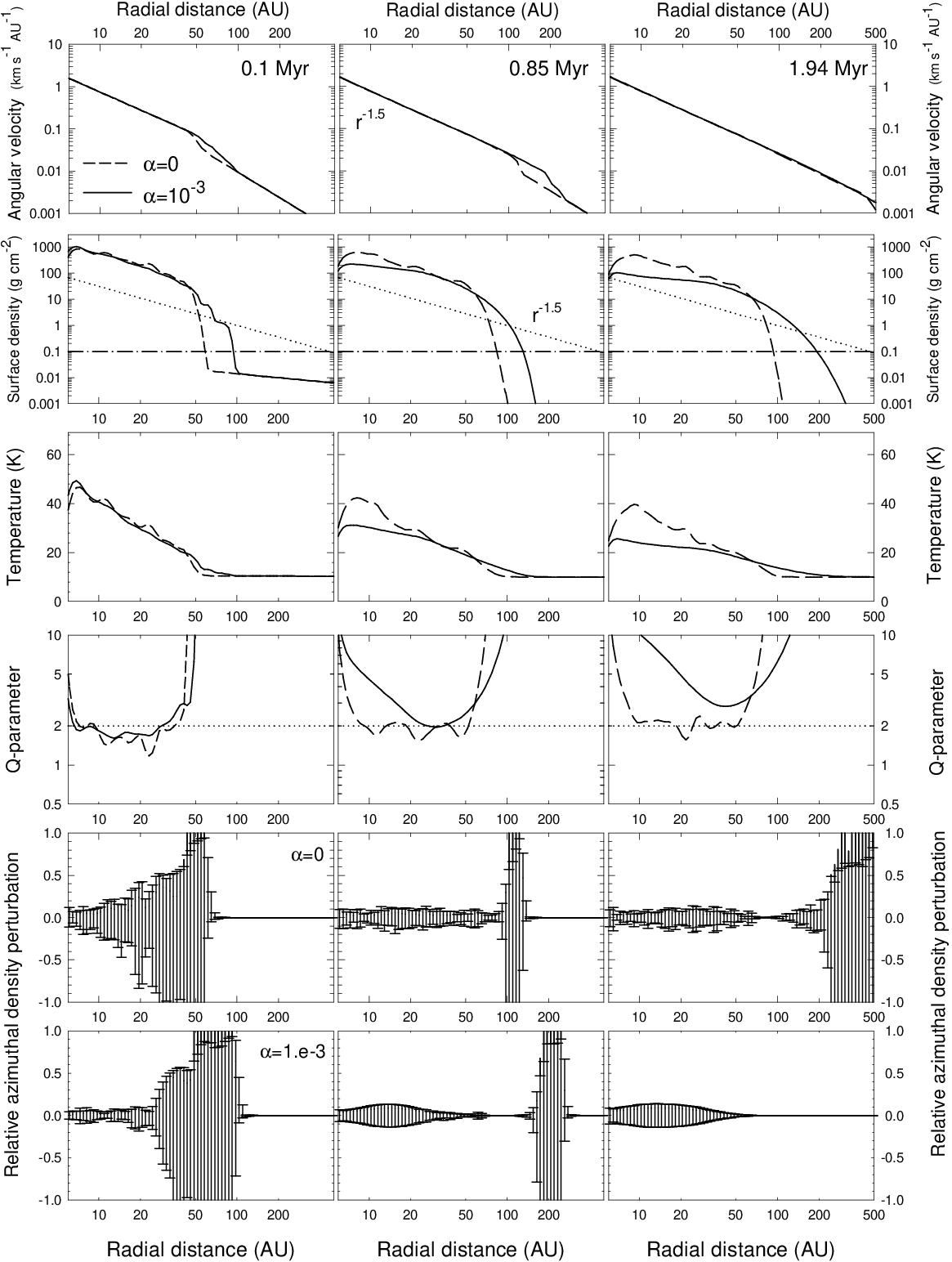}}
      \caption{Radial structure of the circumstellar disc in the $\alpha=0$ model~1 (dashed lines) 
      and $\alpha=10^{-3}$ model~3 (solid lines). See captions to Fig.~\ref{fig1} for details.}
         \label{fig2}
\end{figure}

\subsection{Radial profiles}
\begin{figure}
  \resizebox{\hsize}{!}{\includegraphics{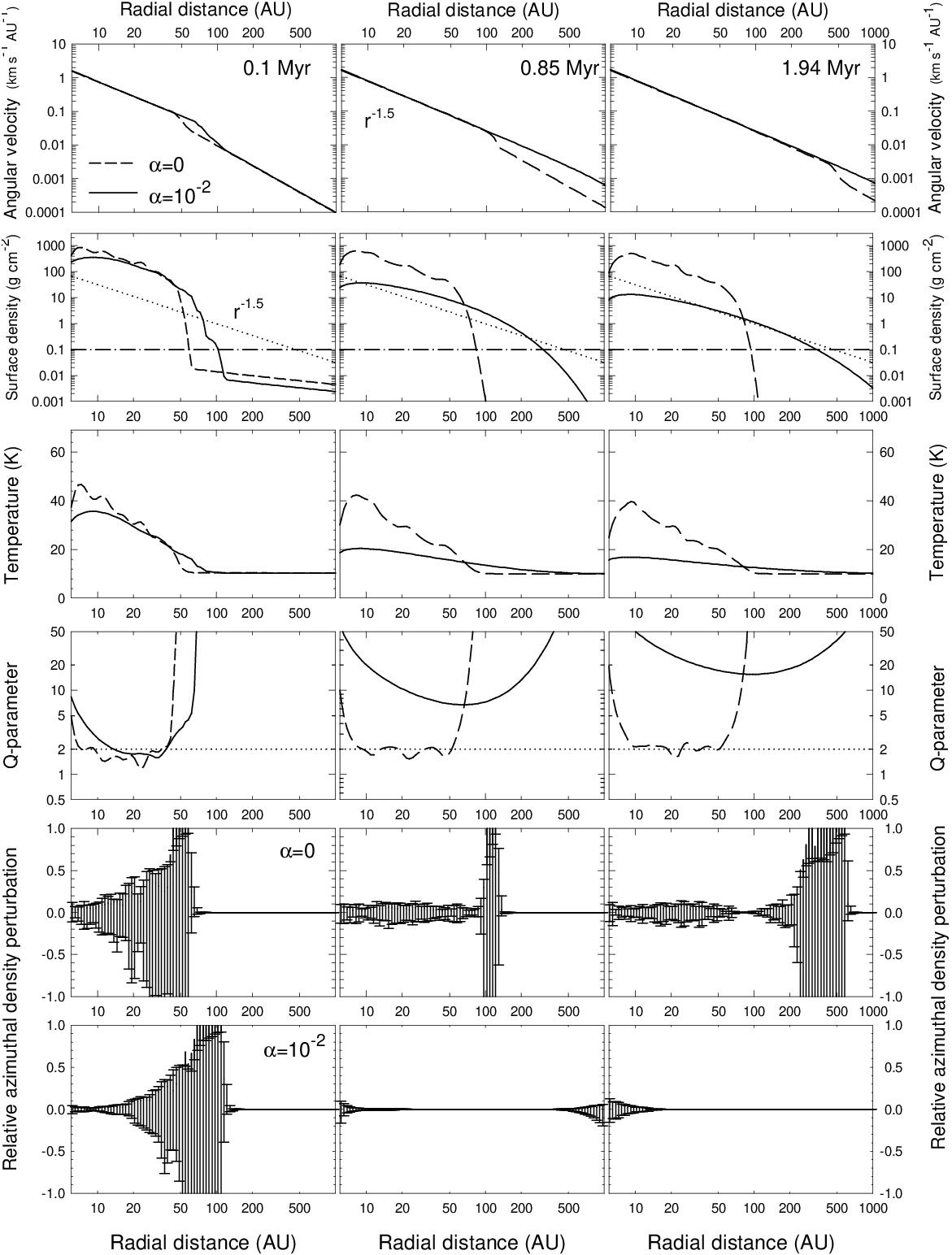}}
      \caption{Radial structure of the circumstellar disc in the $\alpha=0$ model~1 (dashed lines) 
      and $\alpha=10^{-2}$ model~4 (solid lines). See captions to Fig.~\ref{fig1} for details.}
         \label{fig3}
\end{figure}

\begin{figure}
  \resizebox{\hsize}{!}{\includegraphics{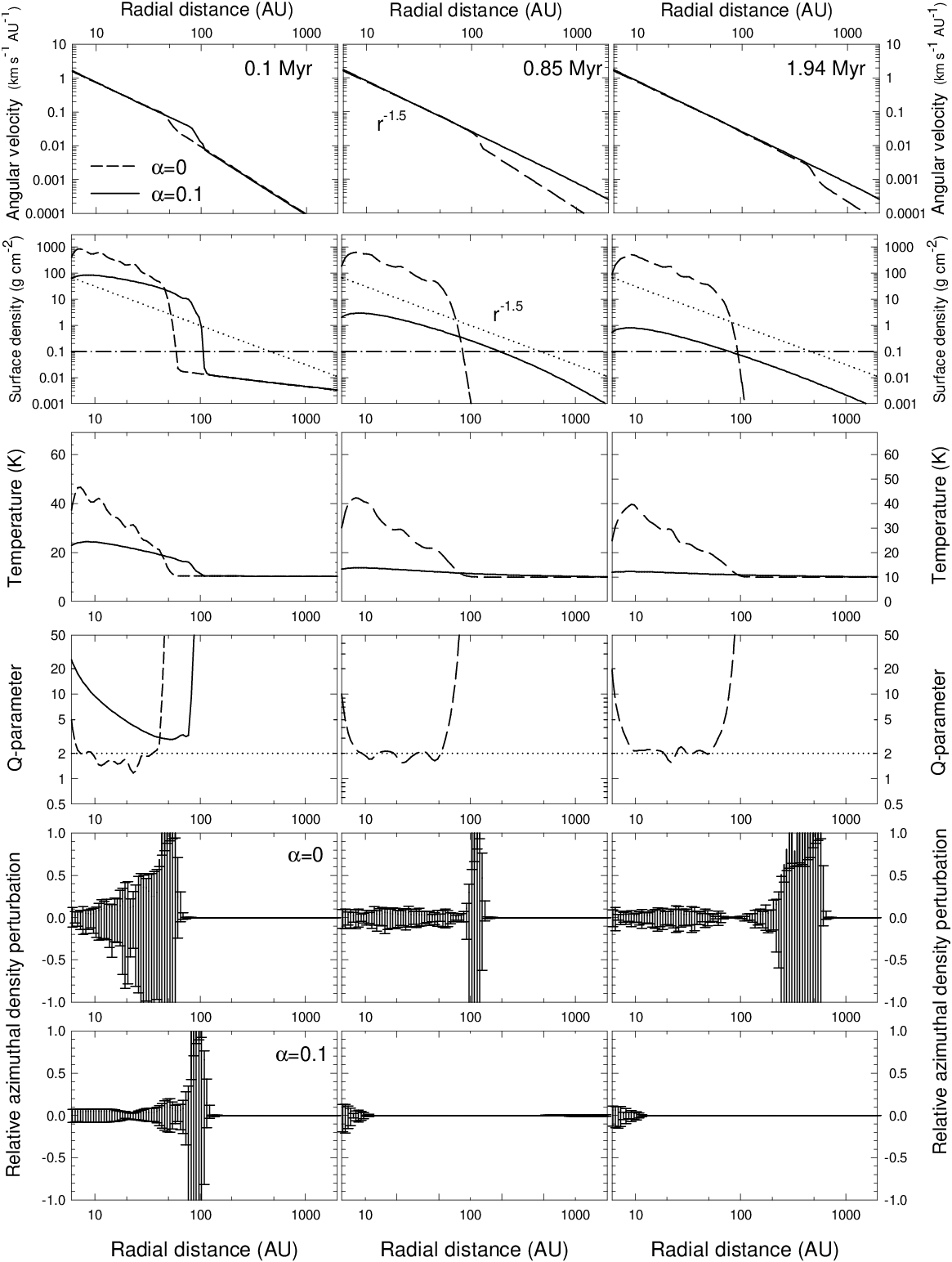}}
      \caption{Radial structure of the circumstellar disc in the $\alpha=0$ model~1 (dashed lines) 
      and $\alpha=10^{-1}$ model~5 (solid lines). See captions to Fig.~\ref{fig1} for details.}
         \label{fig4}
\end{figure}

Solid lines in Figs~\ref{fig1}-\ref{fig4} show three distinct 
snapshots in the evolution of a circumstellar disc in model~2 (Fig.~\ref{fig1}),
model~3 (Fig.~\ref{fig2}), model~4 (Fig.~\ref{fig3}), and model~5 (Fig.~\ref{fig4}).
The numbers in the top horizontal row indicate ages of the disc, $t_{\rm disc}$.
Horizontal rows in each figure show (from top to bottom)
the {\it azimuthally averaged} radial profiles of angular velocity $\overline{\Omega}$, 
surface density $\overline{\Sigma}$, temperature $\overline{T}$, Toomre $\overline{Q}$-parameter, 
and the relative surface density perturbation $\Delta \Sigma$.
Dashed lines in each figure give the corresponding radial profiles for the $\alpha=0$ 
model~1.
The Toomre parameter is calculated as $Q =\tilde{c}_s \Omega/(\pi G \Sigma)$.
In each computational zone ($r_i,\phi_j$) we calculate the relative azimuthal perturbation
to the surface density
\begin{equation}
\Delta \Sigma(r_i,\phi_j) = {\Sigma(r_i,\phi_j)  - {1\over N} \sum_{j=1}^N \Sigma(r_i,\phi_j)
\over \Sigma(r_i,\phi_j), }
\end{equation}
where $N$ is the number of grid zones in the azimuthal direction. The error bars in two bottom 
rows of Figs~\ref{fig1}-\ref{fig4} show the minimum and maximum 
$\Delta \Sigma(r_i,\phi_i)$ in each radial annulus.

It is evident that $\alpha=10^{-4}$ (Fig.~\ref{fig1}) has little effect on the secular
evolution of a self-gravitating disc. In the end of numerical simulations, 
when the disc is 1.94-Myr-old, 
the radial profiles of the viscous and non-viscous models are nearly identical. 
The maximum disc radius is about 100~AU\footnote{The disc is defined as the radial distance at which 
the surface density drops below 0.1 g~cm$^{-2}$} and both the angular velocity and surface density
scale as $r^{-3/2}$. In the early evolution ($t_{\rm disc}=0.1$~Myr), the disc is prone to 
gravitational 
instability as indicated by both the low values of $Q=1.2-1.7$ and large-amplitude 
azimuthal density perturbations $\Delta \Sigma=0.2-1.0$. In the late evolution 
($t_{\rm disc} \ga 0.85$~Myr), 
the disc regulates itself near the boundary of gravitational stability, $Q=1.7-2.0$.
This state is characterized by ongoing low-amplitude density perturbations
powered by swing amplification at the disc's sharp outer edge, resembling truncated
circumstellar discs seen in young stellar objects.
The dotted lines in the surface density panels in Figures~\ref{fig1}-\ref{fig4} 
show the radial surface density
profile as expected from the minimum mass solar nebula (MMSN) model, 
$\Sigma_{\rm mmsn}=10^{3} \, (r/{\rm AU})^{-3/2}$~g~cm$^{-2}$ \citep{WSh77}. Our model 
disc is approximately a factor of ten more dense than the MMSN. 
This implies that most of the disc is actually in the optically {\it thick} regime,
which may have important consequences for the disc mass measurements in 
T Tauri stars \citep{Andrews}.

As the magnitude of $\alpha$ is increased to $10^{-3}$ (Fig.~\ref{fig2}),
numerical simulations start to show noticeable differences between viscous 
and non-viscous discs in the late evolution ($t_{\rm disc} \ga 0.85$~Myr). 
In particular, the disc starts to spread out
and its sharp outer edge is replaced with a shallow tail. The disc radius 
at $t_{\rm disc}=1.94$~Myr is twice as large in model~3 ($\approx 200$~AU) as that 
in the non-viscous model~1. Both $\Sigma$ and $T$ decrease throughout the disc.
In contrast to non-viscous discs, the radial surface density profile in model~3 
cannot be fitted by a single slope. In particular, $\Sigma$ scales as $r^{-0.8}$
in the radial range $5-60$~AU but becomes progressively steeper in the outer portion
of the disc. It is also worth noting that viscous discs start to slowly drift apart
from a self-regulation state characterized by a near constant $Q\approx 1.7-2.0$. 
For example, the minimum $Q$ value in model~3 at $t_{\rm disc}=1.94$~Myr is $2.8$.  

As we continue to increase the value of $\alpha$ to $10^{-2}$ (Fig.~\ref{fig3}), 
the long-term effect of viscosity on the disc evolution becomes more profound. 
It is only the early phases of evolution in model~4 ($t_{\rm disc}=0.1$~Myr) that bear some 
similarities with model~1, the late evolution is considerably different.
Figure \ref{fig3} indicates an overall 
decrease in the values of $\Sigma$ as compared to those in the non-viscous model~1, which 
results in a disc that is optically thin and cold at $t_{\rm disc} \ga 0.85$~Myr. 
The values of $\Sigma$ in the radial range $10-200$~AU are similar, within a factor of few,
to those of the MMSN (dotted lines) but become lower in the entire disc after 1.0~Myr.
It is important to note that planet formation models \citep[e.g.][]{Ida04} 
require discs with gas surface densities a few times greater than that of the MMSN.
The radial surface density profile cannot 
be characterized by a single slope. It is nearly flat
in the inner part of the disc and steepens out in the outer parts.
The disc radius amounts to roughly $350-400$~AU 
at $t_{\rm disc}=1.94$~Myr, though we that note there is no clear 
disc boundary and the surface density gradually declines with radius 
to the values typical for molecular cloud cores.
The disc is virtually {\it axisymmetric} in the late evolution, except for
a small portion near the inner boundary, and is characterized by $Q\gg1.0$.
This marks the largest and most noticeable difference between the purely self-gravitating 
disc and the one with $\alpha=0.01$ -- the latter is profoundly gravitationally stable.

The gravitational stabilization of viscous discs along a sequence of increasing
$\alpha$ is accompanied by progressively more axisymmetric structure. This
is an important point to keep in mind given the available
observational data demonstrating that 1-Myr-old discs are 
non-axisymmetric and show elements of spiral arms and arcs
\citep[e.g.][]{Fukagawa,Grady}\footnote{Given a large size and relatively low mass 
of the disc observed in AB~Aurigae, the nonaxisymmetry in this case may be triggered by 
some kind of close encounter.}. 
Furthermore, we point out that direct gravitational instability as a
scenario for giant planet formation is not viable for the 
viscous-dominated models. 

An increase in $\alpha$ to $10^{-1}$ (Fig.~\ref{fig4}) has a catastrophic effect 
on the secular evolution of a circumstellar disc. After two million years of evolution,
the disc is virtually washed out as a result of a very efficient mass and angular momentum 
radial transport. In particular, the values of $\Sigma$ are more than an order 
of magnitude smaller than those of the MMSN at all radii. The Toomre $Q$-parameter
is much larger than unity and is off the scale in Fig.~\ref{fig4} at $t\ga 0.85$~Myr.
Definitely, circumstellar discs cannot sustain such large values of $\alpha$ 
for a long time without being completely destroyed.

Figs~\ref{fig1}-\ref{fig4} show that the angular velocity of the inner disc 
increases with time only by a few per cent. One may expect this trend be
more profound as the star accumulates mass from the disc.
However, by the time disc forms in our numerical simulations ($\sim 0.14$~Myr),
the star has already gained most of its final mass (see Fig.~\ref{fig6}). 
This is partly caused by a relatively low initial rate of rotation of our cloud core
and partly by the use of a finite-size (5~AU) sink cell in our code. We cannot resolve
the very early phases of disc evolution.

\subsection{Gravitational torques versus viscous torques}

\begin{figure}
  \resizebox{\hsize}{!}{\includegraphics{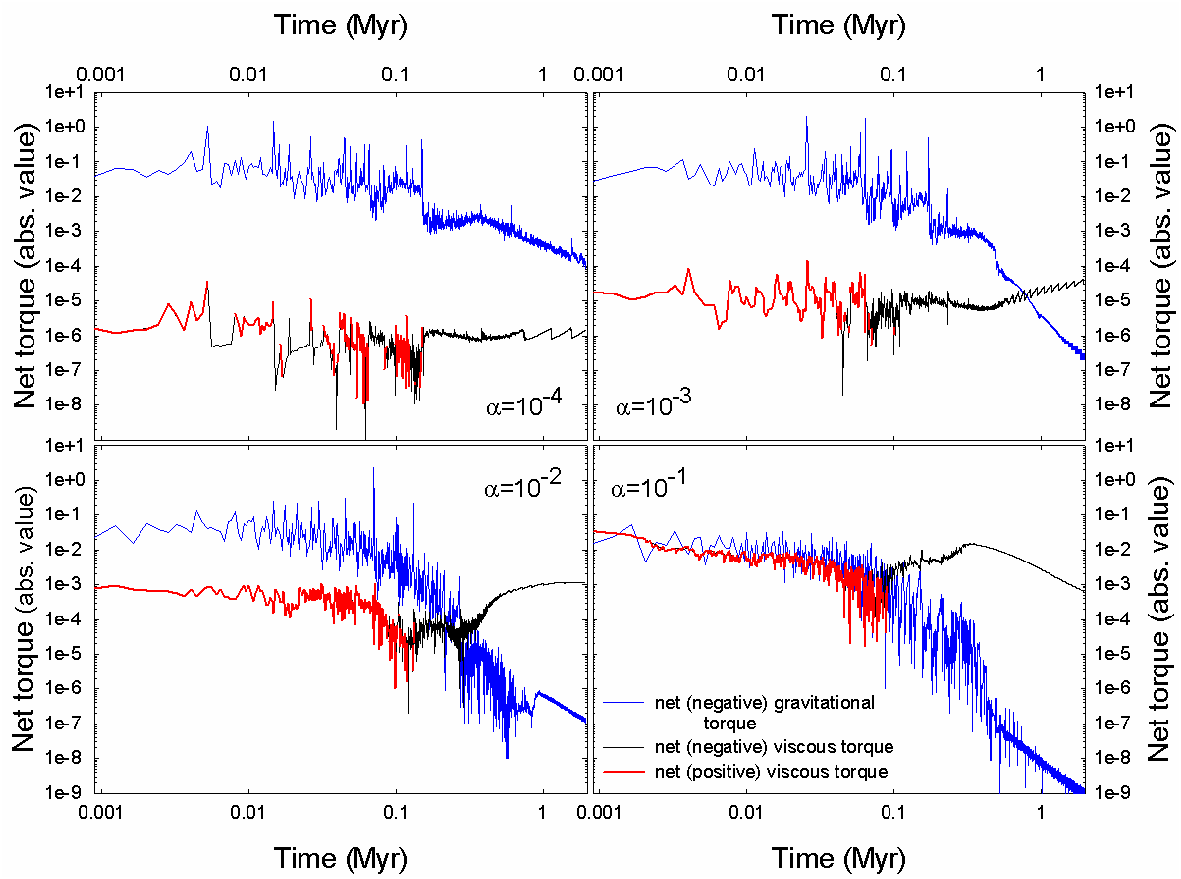}}
      \caption{Temporal evolution of gravitational and viscous net
      torques (in absolute values) in model~2 (upper left), 
      model~3 (upper right), model~4 (lower left), and model~5 (lower
      right). The horizontal axis shows time elapsed since the disc formation.
      The net gravitational torque (blue lines) is always negative, 
      while the net viscous torques can be both positive (red lines) and negative (black lines).
      The torque is measured in units of $8.66 \times 10^{40}$~gm~cm$^{2}$~s$^{-2}$. }
         \label{fig5}
\end{figure}

The differences in the secular evolution of models~$1-4$ can be understood if 
we compare the temporal evolution of gravitational and viscous torques in these models.
The blue lines in Figure~\ref{fig5} show the net (positive plus negative) gravitational torque in 
the disc ${\cal T}_{\rm g}$, while the red and black lines show the net viscous 
torque ${\cal T}_{\rm v}$. The horizontal axis shows time elapsed since the disc formation. 
Both the gravitational and viscous torques are plotted 
in {\it absolute} units and the net gravitational torque is always {\it negative}. 
However, the net viscous torque is found to change its sign.
In the early disc evolution at $t \la 0.1$~Myr, the net viscous torque is mostly 
positive and its magnitude is plotted with red. In the subsequent evolution,
the net viscous torque becomes negative and its magnitude is plotted with black.
The adopted values of $\alpha$ are indicated in each panel.
The net gravitational torque ${\cal T}_{\rm g}$ is found by summing 
all local gravitational torques $\tau_{\rm g}(r,\phi) = - m(r,\phi)\, \partial 
\Phi / \partial \phi$  in each computational cell occupied by the disc, 
where $\Phi$ is the local gravitational 
potential and $m(r,\phi)$ is the gas mass in a cell with polar coordinates ($r,\phi$).
The net viscous torque is found in a similar manner by summing up all 
local viscous torques
$\tau_{\rm v}(r,\phi)=r\, (\nabla \cdot \mathbf{\Pi})_\phi  \, S(r,\phi) $, where $S(r,\phi)$
is the surface area occupied by a cell with polar coordinates ($r,\phi$).
The local torques $\tau_{g}(r,\phi)$ and $\tau_{\rm v}(r,\phi)$ are actually the local 
azimuthal components of the corresponding 
forces, acting on a fluid element with mass $m(r,\phi)$, multiplied by the radius $r$.

Figure~\ref{fig5} clearly demonstrates that gravitational torques dominate over viscous 
torques in the $\alpha=10^{-4}$ model. The difference is particularly large in
the early evolution and is decreasing later but ${\cal T}_{\rm g}$ remains at least an order 
greater than ${\cal T}_{\rm v}$ in the 2-Myr-old disc. A qualitative change in the temporal 
behaviour of torques is seen in the $\alpha=10^{-3}$ model -- while the early disc evolution 
($t \la 0.8$~Myr) is still controlled by gravity, the subsequent evolution is viscosity dominated.
This effect is even more prominent in the $\alpha=10^{-2}$ disc, in which the viscous torques start to
prevail already after 0.3~Myr of evolution. This explains profound changes in the disc structure 
seen in Fig.~\ref{fig3}.
The $\alpha=10^{-1}$ model is an extreme case, in which viscous torques compete with those of gravity
even in the early disc evolution and completely prevail in the late evolution.

It is important to note here that gravitational stability properties 
of astrophysical discs may be modified in the presence of magnetic fields. For instance, 
\citet{Fromang04a,Fromang04b} have demonstrated that the MRI leads to turbulence, 
which gives rise to a complicated spiral pattern in the disc and lowers
the strength of the gravitational stress tensor by a factor of 2 due to the nonlinear 
mode-mode interaction.  In addition, frozen-in magnetic 
fields tend to decrease the critical Toomre parameter $Q_{\rm cr}$ 
and the magnetic disc has to attain a lower value of $Q$ in order to destabilize \citep{VB2}.
Since $\alpha$-viscosity is most likely caused by the MRI, we acknowledge that the actual 
strength of gravitational torques may be somewhat smaller than that shown in Fig.~\ref{fig5}. 
We plan to investigate the effect of magnetic fields and magnetic braking in a follow-up paper. 

Two important features in Figure~\ref{fig5} need to be emphasized. 
First, a decline of gravitational torques with time is
mostly caused by a gradual approach of the disc to a stable state. 
As a consequence, large scale spiral arms
in the early phase are replaced with ongoing low-amplitude density perturbations in the late phase 
\citep{VB3}. As Figure~\ref{fig5} demonstrates, turbulent viscosity expedites the onset of gravitational
stability and the $\alpha=10^{-2}-10^{-1}$ discs are virtually axisymmetric and gravitationally stable
after just 1.0~Myr of evolution. 
Second, a sum of net viscous and gravitational torques is always {\it negative}.
This fact constitutes a fundamental property of self-gravitating circumstellar discs, both viscous
and non-viscous.  
The net negative torque is related to the ongoing accretion of gas
on to a protostar -- the accreted material removes the disc angular momentum, which may be later
ejected to the external environment via magnetic braking and protostellar jets. As a consequence, 
the rate of change of the disc angular momentum and the net torque are both negative. 
It is also interesting to note that the net viscous torque in the first 
$10^5$~yr of evolution is positive. This is related to the fact that most of the disc in this phase
is characterized by the dynamic viscosity that declines with radius faster than $r^{-0.5}$.
Indeed, it can be shown that an axisymmetric Keplerian disc has a positive viscous torque, i.e.
$r \left(\nabla \cdot\Pi \right)_\phi>0$ if $\mu \propto r^{\beta}$, where $\beta<-0.5$.

\subsection{Accretion rates and disc masses}
\label{diskmass}
Figure~\ref{fig5} demonstrates that circumstellar discs evolve through two distinct physical regimes
during their long-term evolution, gravity-dominated and viscosity-dominated.  
The former always precedes the latter and the time when the viscosity-dominated regime takes
over, or even its existence, depends on the value of $\alpha$. In each of the two regimes, 
radial transport of mass and angular 
momentum is controlled by principally different mechanisms and 
the mass accretion rates on to the central star should bear the imprints of these differences.

\begin{figure}
  \resizebox{\hsize}{!}{\includegraphics{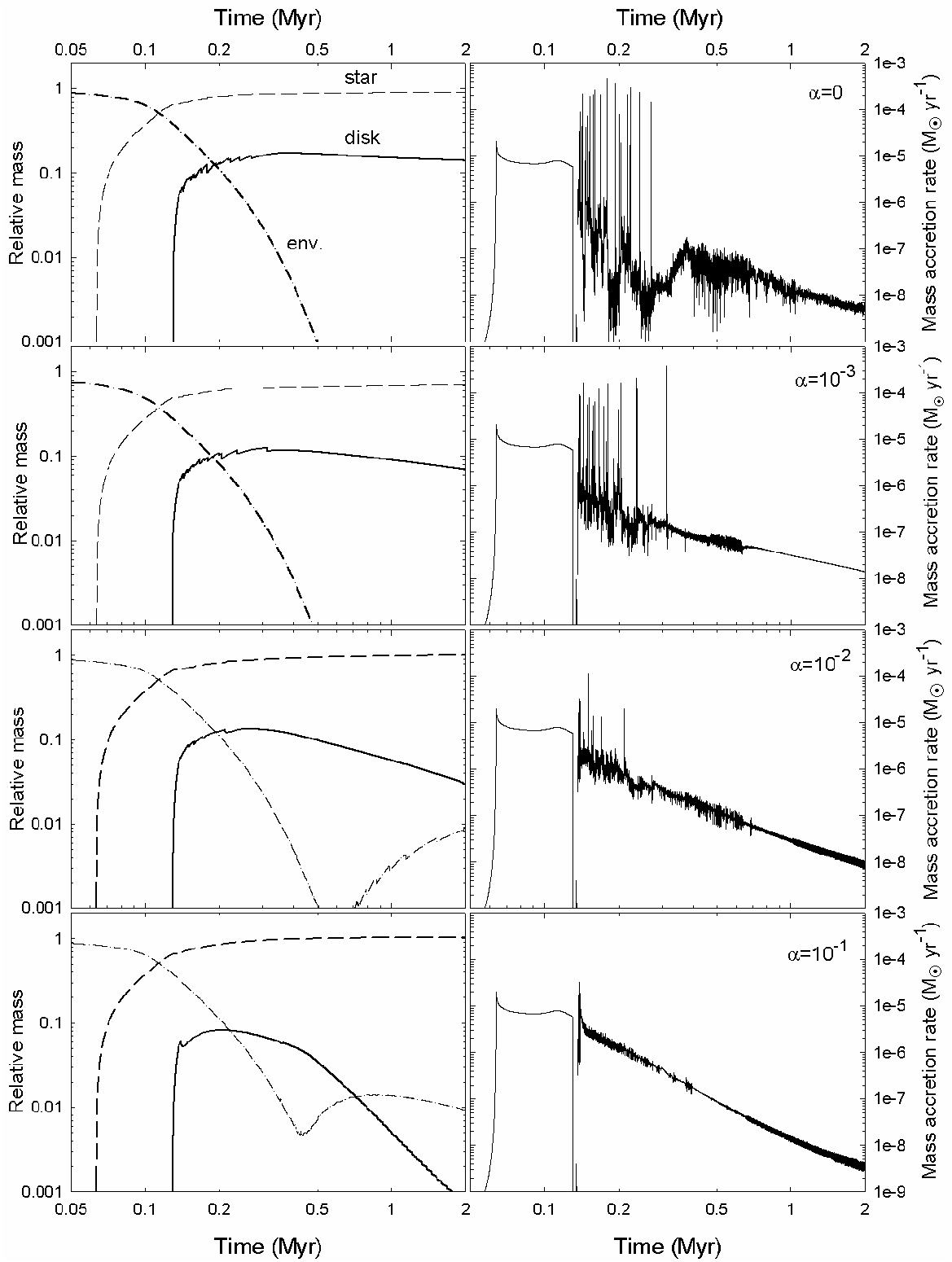}}
      \caption{Right column: temporal evolution of the mass accretion rate $\dot{M}$ in the 
      $\alpha=0$ model~1 (top), $\alpha=10^{-2}$ model~4 (middle), and $\alpha=10^{-1}$ model~5 (bottom).
      Left column: temporal evolution of the disc mass (solid lines), stellar mass (dashed lines), 
      and envelope mass (dot-dashed lines) in model~1 (top), model~4 (middle), and model~5 (bottom).
      All masses are relative to the initial cloud mass $M_{\rm cl}=0.8~M_\odot$.}
         \label{fig6}
\end{figure}

The right column in Fig.~\ref{fig6} shows (from top to bottom) 
the mass accretion rate in model~1 ($\alpha=0$), model~3 ($\alpha=10^{-3}$), model~4 ($\alpha=10^{-2}$),
and model~5 ($\alpha=10^{-1}$) as a function of time. 
The mass accretion rate is computed as $\dot{M}=-2 \pi r
\, v_r \Sigma(r)$, where $v_r$ is the radial velocity at the inner inflow boundary $r=5$~AU.
In all four models the evolution starts with a sharp rise of $\dot M$ to a maximum value, 
manifesting the formation of a central star at $t\approx 0.06$~Myr, 
and continues with a short phase of near constant accretion, in which matter is directly accreted 
on to the star.
The disc forms at $t\approx 0.14$~Myr and the subsequent accretion history is considerably 
different between the models. 
The non-viscous model~1 develops short-lived mass accretion bursts\footnote{The burst phenomenon 
is caused by disc fragmentation due to high rates of
mass infall from the envelope in the early embedded phase 
and is described in detail in \citet{VB1,VB2}. A highly variable accretion was also reported in the
case of isolated massive discs by \citet{Lodato05}.  }
with $\dot{M}\ge 10^{-4}~M_\odot$~yr$^{-1}$, while the quiescent phase is characterized
by a much lower accretion rate in the range ${\rm a~few} \times 10^{-7}-10^{-8}~M_\odot$~yr$^{-1}$.
On the other hand, the $\alpha=10^{-1}$ model~5 shows virtually no bursts, while 
the $\alpha=10^{-2}$ model~4 has only a few of them. It is evident that the burst activity 
diminishes along a sequence of increasing $\alpha$.
There is {\it no} quiescent phase of low-rate accretion in models~4 and 5 --
the mass accretion rate gradually declines with time from a few $\times 10^{-6}~M_\odot$~yr$^{-1}$ 
just after the disc formation to $10^{-8}~M_\odot$~yr$^{-1}$ (and lower) at 2.0~Myr. 

There are two observationally valuable consequences of the burst phenomenon.
First, it can provide an explanation for the FU Orionis variables. The apparent lack of 
bursts in the viscous discs will make it more difficult to account for the burst activity, 
though other burst mechanisms may be operational in the inner disc at $r \la 1$~AU 
\citep[e.g.][]{Bell94}.
Second, the existence of the quiescent phase of accretion 
in the {\it early} evolutionary phase can potentially be linked with a recent 
detection of Very Low Luminosity Objects (VeLLOs) by the {\it Spitzer Space Telescope} 
\citep[e.g.][]{Young}. A previously detected object in this category is IRAM 04191 \citep{Andre}; 
for another example see \citet{Stecklum}.
All of these are objects with luminosity 
$L\le 0.1 L_\odot$ embedded within dense cores, which, based on previous observations,
were often classified as starless. Their association with dense cores
and their low luminosity suggests that they are {\it young} objects   
that feature some combination of a sub-solar mass and low accretion rate.
They might be proto-brown dwarfs, but Fig.~\ref{fig6} offers an alternative.
The non-viscous model~1 can account for VeLLOs as young protostars in the quiescent 
accretion phase with
$\dot{M}\sim 10^{-8}~M_\odot$~yr$^{-1}$, which for the protostellar mass of $\sim 1.0~M_\odot$
would correspond to the accretion luminosity $3\%$ that of the solar.
Clearly, the viscous models~4 and 5 cannot account for VeLLOs, as their mass accretion rates
in the early disc evolution are too high. 
The identity of these objects can ultimately be revealed  by studying the 
chemical evolution of brown dwarfs and young protostars \citep{Lee}.

Distinct changes in the radial structure of circumstellar discs along the sequence
of increasing $\alpha$, as seen in Figs~\ref{fig1}-\ref{fig4}, suggest that 
viscosity and gravity-dominated discs may have quite different masses.
The left column in Fig.~\ref{fig6} shows disc masses (solid lines), stellar masses (dashed
lines) and envelope masses (dot-dashed lines), all relative to the initial cloud core mass $M_{\rm cl}=0.8~M_\odot$,
in (from top to bottom) model~1, model~3, model~4, and model~5. In the $\alpha=0$ model~1 
the relative disc mass reaches a maximum value of $14\%$ soon after the disc formation and is not 
significantly changing afterwards. On the contrary, disc masses in the $\alpha=10^{-2}$
models~4 and $\alpha=10^{-1}$ model~5, though attaining similar maximum values soon after
the disc formation, show a significant 
decline with time in the subsequent evolution. In particular,
the disc mass in model~4 drops to only $3\%$ that of the central star at 2~Myr, while 
the disc in model~5 is virtually depleted after 2~Myr of evolution. 
At a first glance, it may seem surprising that viscous and non-viscous models have 
similar (within a factor of 2) disc masses in the early evolution although 
the mass accretion rates in the viscous models appear (by eye) to be greater than those
in the non-viscous model. This paradox is resolved by the fact that the bursts are very efficient 
in regulating the disc mass. During each burst a significant amount of mass ($0.01-0.05~M_\odot$) 
is accreted on to the protostar \citep{VB1,VB2}. 

A decline in the disc mass in models~4 and 5 seen in the late evolution is caused not only by
accretion on to the central star but also by disc expansion due to viscous torques. 
The dot-dashed lines in Fig.~\ref{fig6}
illustrate this phenomenon -- the envelopes in models~4 and 5 are not completely depleted
by the end of numerical simulations (as in model~1 and 3) but appear to gain some mass 
in the late evolution. This mass is coming from the disc, which dilutes and expands 
in the late evolution (see Figs~\ref{fig3} and \ref{fig4}), 
thus losing part of its material to the external environment. 
The disc expansion is explained by the fact that $\mu$ in the outer disc scales as $r^{\beta}$,
where $\beta<-0.5$, which causes the disc material to be transported outward 

Figure~\ref{fig6} demonstrates an important ingredient of our numerical model -- our discs
are formed self-consistently during numerical simulation and not introduced in the beginning 
of numerical simulations. Indeed, the time evolution of the mass accretion rate 
(MAR) in Fig.~\ref{fig6} can be split in to three phases. In the first 
phase, the MAR is characterized by a sharp growth from a negligible value to a 
peak value of approximately $2\times 10^{-5} M_\odot$~yr$^{-1}$. 
This behaviour is characteristic 
for the runaway collapse phase and stellar core formation phase in spherically 
symmetric  cloud core collapse simulations \citep[e.g.][]{Foster,VB5}. 
The second phase is characterized by a near-constant MAR 
at approximately $10^{-5}~M_\odot$~yr$^{-1}$. This is typical for collapsing self-similar 
spherically symmetric cloud cores \citep[e.g.][]{Shu77}. In these two early phases, 
the infalling envelope accretes directly onto the central star. 
At approximately 0.14~Myr a first 
infalling layer of gas hits the centrifugal barrier at $r=5$~AU and a centrifugally 
balanced disc begins to form\footnote{In reality, the disc forms earlier but we cannot resolve
its evolution within the inner 5~AU due to the use of the sink cell}. Since then, the system 
enters a qualitatively distinct phase of disc accretion. 
Because the forming disc is too small and in the state of near centrifugal balance, 
the MAR drops by many orders of magnitude. 
This sharp drop is a signature of the disc formation and cannot be present in 
numerical simulations if they start right from the disc phase. 
Soon after the disc has formed, it accumulates enough mass to trigger 
radial mass transport and the subsequent time behaviour of the MAR is controlled
by the interplay of self-gravity and viscosity. 

\section{Stiffer equation of state}
\label{hot}
Numerical simulations with a more realistic treatment of
the energy balance indicate that the strength of gravitational instability in general 
and the disc propensity to
fragmentation in particular depend on the characteristic time of disc cooling and, hence, 
on the disc temperature \citep[e.g.][]{Gammie01,Johnson03,Rice03,Lodato04, Mejia05}.
In this section we consider the effect of increasing the ratio of specific heats from 
$\gamma=1.4$ to $\gamma=1.67$ in a purely self-gravitating disc. 
This stiffening of the barotropic equation of state 
should raise disc temperatures and mimic the effect of a longer characteristic cooling 
time.

\begin{figure}
  \resizebox{\hsize}{!}{\includegraphics{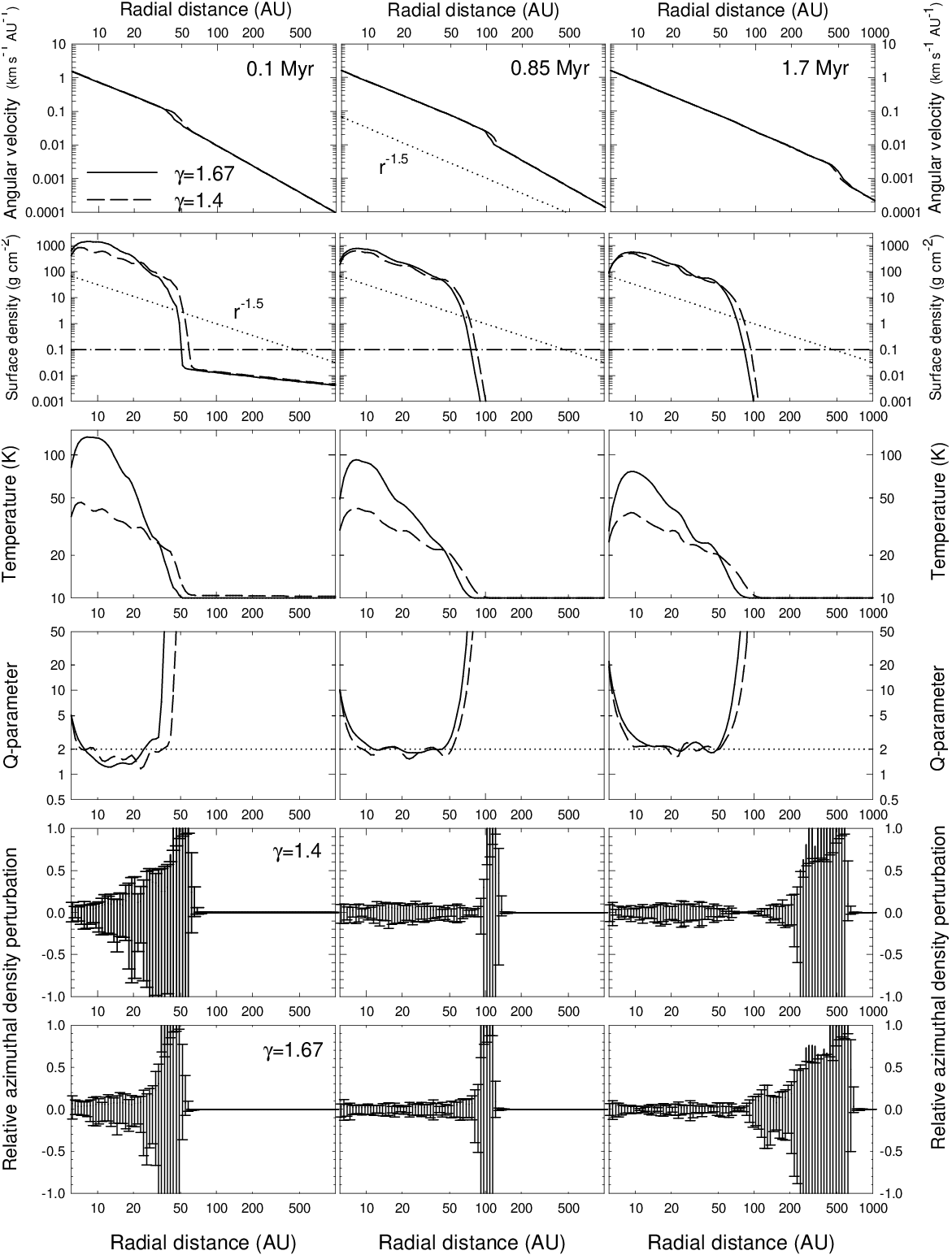}}
      \caption{Radial structure of a purely self-gravitating disc ($\alpha=0$) 
      in the $\gamma=1.4$ model (dashed lines) and $\gamma=1.67$ model (solid lines). The disc age is
      indicated in the top frame of each column.  See captions to Fig.~\ref{fig1} for details.}
         \label{fig7}
\end{figure}

In Fig.~\ref{fig7} we compare the disc radial structure 
in models with $\gamma=1.4$ (dashed lines)
and $\gamma=1.67$ (solid lines). It is evident that an increase in $\gamma$ results in a factor
of two higher gas temperatures in the disc, especially in its inner region. At the same time, the
gas surface density also increases and the resulting Toomre $Q$-parameter
changes insignificantly and (in some parts of the disc) 
may even drop below the values characteristic for the colder disc. 
The radial distribution of the gas surface density scales as $r^{-1.5}$
throughout most of the evolution, except for the very early phase when 
$\Sigma\propto r^{-2.0}$. The relative azimuthal density perturbations are
slightly smaller than those in the colder disc. To summarize, the most substantial change is 
seen in the radial gas temperature distribution. However, since the sound speed is proportional 
to the square root of the gas temperature and the gas surface density also  increases, 
the resulting Toomre parameter does not change significantly.

\begin{figure}
  \resizebox{\hsize}{!}{\includegraphics{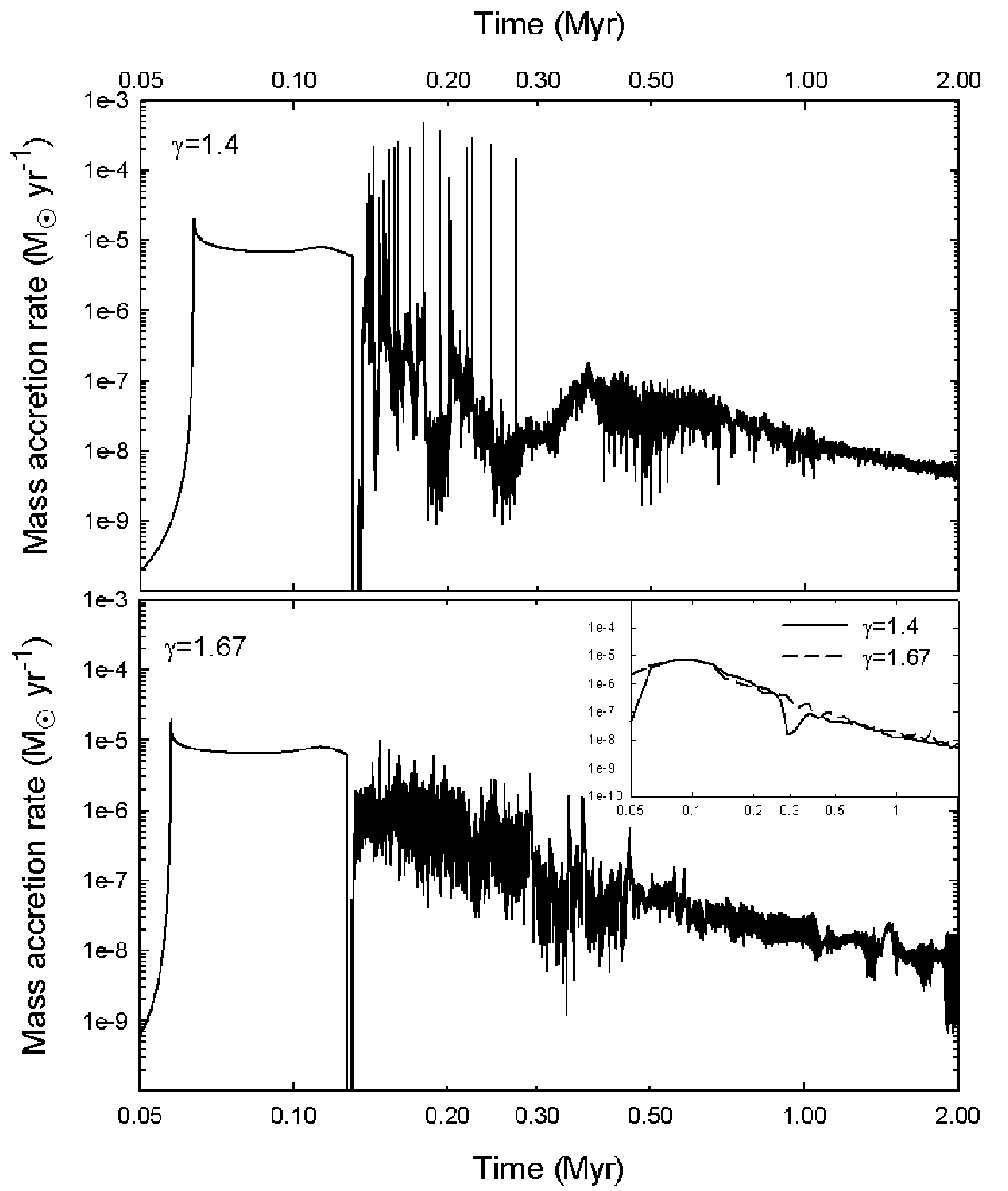}}
      \caption{Mass accretion rates in the $\gamma=1.4$ model (top) and $\gamma=1.67$ model (bottom)
      as a function of time. The both models are non-viscous ($\alpha=0$). The insert shows the
      rates averaged over a period of $2\times 10^4$~yr in the $\gamma=1.4$ model (dashed line) and
      $\gamma=1.67$ model (solid line).}
         \label{fig8}
\end{figure}

Our next step is to compare the mass accretion rates in models with 
different values of $\gamma$. The top and bottom panels in Fig.~\ref{fig8} show $\dot{M}$ 
in models with $\gamma=1.4$ and $\gamma=1.67$, respectively. A visual inspection of the figure
reveals that the most noticeable change in the time behaviour of $\dot{M}$ occurs
in the early phase of disc evolution between $0.14$~Myr and $0.4$~Myr. While the $\gamma=1.4$
disc exhibits extremely varying rates with short bursts followed by 
longer periods of quiescent accretion, the $\gamma=1.67$ disc shows a gradually declining 
accretion rate with an order of magnitude fluctuations or flickering around mean values.
However, the time-averaged (over $2\times 10^4$~Myr) mass accretion rates  are not too dissimilar,
as is seen in the insert to Fig.~\ref{fig7}. More specifically, the $\gamma=1.4$ disc is 
characterized by slightly larger/smaller accretion in the early/late phase than 
the $\gamma=1.67$ disc. It is worth mentioning that this tendency is also seen when 
shorter characteristic cooling times are considered. For instance, 
the $\gamma=1.2$ disc\footnote{It is problematic to consider values of $\gamma$ smaller than $\gamma=1.2$,
since equation~(\ref{barotropic}) tends to overestimate gas pressure when $\gamma \rightarrow 1.0$.} has slightly larger/smaller accretion in the early/late phase than 
the $\gamma=1.4$ disc. In any case, accretion rates show only a factor of unity difference.

Why are the time-averaged mass accretion rates rather similar in both models? 
At a first glance, one might expect a much large contrast, since an increase in gas 
temperature is expected to moderate gravitational instability and 
reduce accretion triggered by gravitational torques. However, as was noticed by \citet{Cai08},
a moderate increase in disc temperature may in fact promote accretion due to the growing 
relative strength of lower order spiral modes in the disc. The global nature of
lower order modes makes them more efficient agents 
for the radial mass transport in the disc than higher order modes, the latter tend to 
produce more fluctuations and cancellation in the net gravitational torque.

\begin{figure}
  \resizebox{\hsize}{!}{\includegraphics{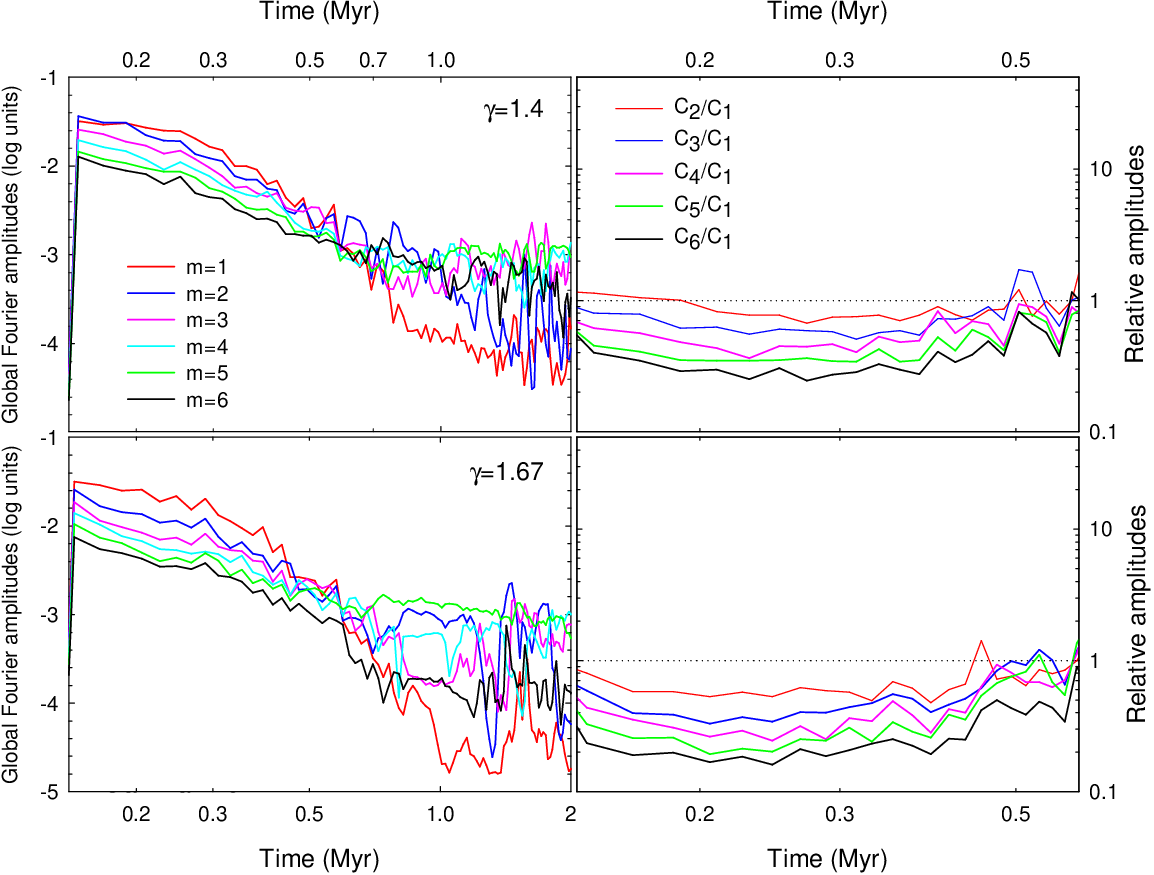}}
      \caption{{\bf Left}: Global Fourier amplitudes $C_{\rm m}$ (in log units) for modes $m=1 - 6$
      as a function of time in the $\gamma=1.4$
      disc (top) and $\gamma=1.67$ disc (bottom). {\bf Right:} Ratio $C_{\rm m}/C_1$
      of the higher order modes $m \ge 2$ to the lowest order $m=1$ mode for the $\gamma=1.4$
      disc (top) and $\gamma=1.67$ disc (bottom). }
         \label{fig9}
\end{figure}

In order to visualize this effect and the strength of spiral modes in the disc, 
we compute the global Fourier amplitudes (GFA) defined as
\begin{equation}
C_{\rm m} (t) = {1 \over M_{\rm d}} \left| \int_0^{2 \pi} \int_{r_{\rm in}}^{r_{\rm disc}} 
\Sigma(r,\phi,t) \, e^{im\phi} r \, dr\,  d\phi \right|,
\end{equation}
where $M_{\rm d}$ is the disc mass and $r_{\rm disc}$ is the disc's physical outer radius.
The instantaneous GFA show considerable fluctuations and we have to time average 
them over $2\times10^4$~yr in order to produce a smooth output.
The temporal evolution of the time-averaged GFA (log units) 
for the $\gamma=1.4$ (top) and $\gamma=1.67$ (bottom) models 
is shown in the left column of Fig.~\ref{fig9}. The time behaviour of the GFA is indicative
of two qualitatively different phases in the disc evolution. In the early phase ($t\la 0.6$~Myr),
a clear segregation between the modes is evident -- the lower order mode dominates its 
immediate higher order neighbour in both models. In particular, the $m=1$ mode is almost always 
the strongest one. The modes also show a clear tendency to decrease in magnitude with time,
which explains a gradual decline in the time-averaged mass accretion 
rates shown in the insert of Fig~\ref{fig8}.
The GFA of the $\gamma=1.4$ disc are somewhat 
larger than those of the $\gamma=1.67$ disc, except probably for $C_1(t)$, 
indicating that gravitational instability is stronger in the $\gamma=1.4$
disc. The same effect is also seen when we compare the $\gamma=1.2$ and $\gamma=1.4$ discs -- 
the former has larger GFA then the latter.
In the late phase, however, this clear picture breaks into a kaleidoscope of modes, with
higher order modes $m\ge 2$ competing for dominance with each other and the $m =1$
mode being almost always the weakest one. The GFA have dropped in magnitude considerably.
However, accretion on to the star does not terminate since the effect of these mode
fluctuations is to produce a net {\it negative} torque in the inner disc \citep{VB3}.
These mode fluctuations are not a numerical noise but are rather 
caused by ongoing low-amplitude
non-axisymmetric density perturbations sustained by swing amplification at the disc's sharp outer edge.
As was shown by \citet{VB3}, self-gravity of the disc is essential for these density perturbations to
persist into the late disc evolution. The density perturbations (and associated accretion) quickly 
disappear if self-gravity is switched off.

In the right column of Fig.~\ref{fig9} we plot the ratios $C_{\rm m}/C_1$ as a function of time  
for the $\gamma=1.4$ disc (top) and $\gamma=1.67$ disc (bottom). 
It is evident that the relative input from the  $m\ge 2$ modes is larger in the $\gamma=1.4$ 
disc than in the $\gamma=1.67$ one.  This implies that the mode-to-mode 
interaction may produce more cancellation in the net gravitational torque in the colder 
$\gamma=1.4$ disc than in the hotter $\gamma=1.67$ disc.  As a result, the efficiency of 
mass transport (due to all modes) reduces in the colder disc and becomes similar to that
of the hotter disc, even though most of the GFA are in fact larger in the former than
in the latter. This may seem counterintuitive but the same effect was found 
in numerical simulations of envelope-irradiated 
discs by \citet{Cai08}. It is worth mentioning that a similar modal behaviour was
found when $\gamma=1.2$ and $\gamma=1.4$ discs are compared in our numerical simulations 
-- the relative input from the  
$m\ge 2$ modes is larger in the $\gamma=1.2$ disc than in the $\gamma=1.4$ one. The effect
is not large but is certainly noticeable.

Alternatively, the similarity of mass accretion rates in discs characterized by different 
$\gamma$ may be merely a result of 
the self-regulating nature of embedded accretion discs, which re-adjust their gravitational torques
(e.g. by increasing/decreasing  surface densities, temperatures, etc.) in order to pass on the mass flux coming from the envelope.

\begin{figure}
  \resizebox{\hsize}{!}{\includegraphics{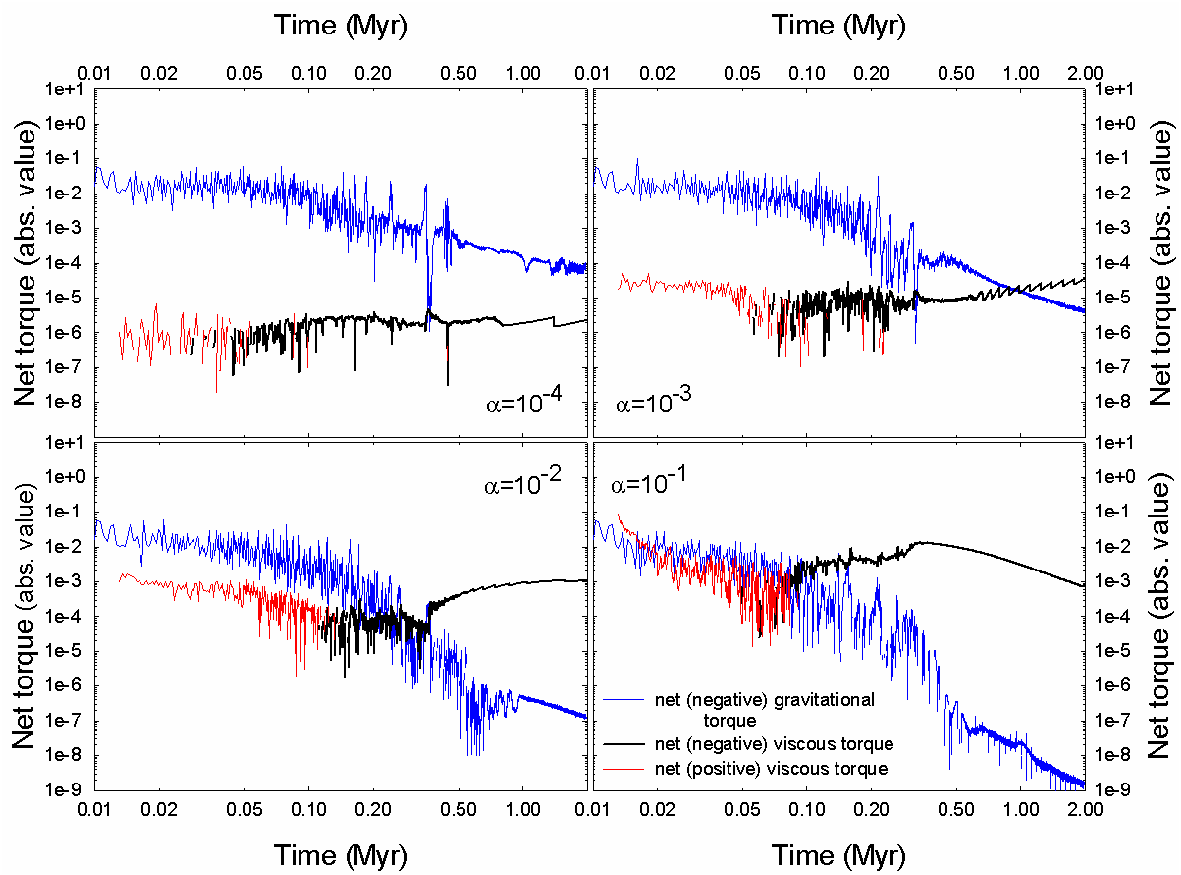}}
      \caption{The same as Fig.~\ref{fig5} only for models with $\gamma=1.67$. }
         \label{fig10}
\end{figure}

Figure~\ref{fig10} shows the temporal evolution of viscous
and gravitational net torques for the same models as in Fig.~\ref{fig5} but with $\gamma=1.67$.   
The comparison of Figs~\ref{fig5} and \ref{fig10} indicates that the time behaviour 
of the torques in both figures is quite similar. There is a slight increase in
the strength of the viscous torques in models with $\gamma=1.67$
caused by a higher (on averaged) disc temperature than in models with $\gamma=1.4$. In particular,
the very early evolution of the $\alpha=0.1$ model ($t \le 0.02$~Myr) is dominated by viscosity, while
in the corresponding model with $\gamma=1.4$ both the viscous and gravitational torques are nearly equal.
We conclude that a (modest) increase in disc temperature does not noticeably affect the disc accretion
properties averaged over many orbital periods ($\sim 10^4$~yr) but can substantially change the 
instantaneous accretion rates.
We also note that although the $\gamma=1.67$ disc does not show vigourous bursts of
mass accretion (but rather an order of magnitude flickering around mean values), the bursts 
can be re-established by increasing the initial rotation velocity of the cloud core 
\citep[e.g.][]{VB2}. 

\section{Discussion}
\label{discuss}

\subsection{Constraints on turbulent viscosity}

There are several known physical processes that can contribute to the radial transport of
mass and angular momentum in circumstellar discs.
These transport mechanisms include gravitational torques, either due to internal (self-gravity) 
or external (companion star) sources, and turbulent viscosity.
The effect of self-gravity on the secular evolution of a circumstellar disc has been considered 
in our previous paper
\citep{VB3}. In this paper we added the effect of turbulent viscosity due to yet unspecified
source of turbulence.
Physical processes in circumstellar discs that can, under specific conditions, give rise to 
turbulence include the magneto-rotational instability and vertical convection. 
The latter may be heavily suppressed in the externally heated discs \citep{Ruden}, which leaves 
the MRI as the most probable source of turbulence in circumstellar discs. 
\cite{Afshordi} have recently argued on analytical grounds that hydrodynamic turbulence 
(not related to the
MRI) can arise in cold Keplerian discs characterized by Reynolds numbers $\ga10^4$
and having both the fine-tuned conditions and appropriate feedback mechanism, 
though this idea has not been proven so far by numerical simulations.

 Many numerical efforts to characterize the MRI-induced turbulent viscosity qualitatively have 
been made in the recent years. These include local shearing box models 
\citep[e.g.][]{Brandenburg96,Stone96,Fleming,Sato04} and global magnetohydrodynamic 
simulations of accretion discs \citep[e.g.][]{Armitage}, 
though the latter are usually limited by a small number of orbital periods.
Most authors calculate  the mean ratios of the Maxwell and Reynolds stresses 
versus midplane gas pressure, $T^{B}_{r\phi}/P_0$ and $T^{Re}_{r\phi}/P_0$, respectively, which, 
to a factor of unity, are proportional to $\alpha$ in Keplerian discs\footnote{In fact, this is true
only for axisymmetric discs, in which $T^{B}_{r\phi}=(d \log\Omega /d \log r) 
\alpha P_0$ and the rate of strain $d \log\Omega /d \log r$ is constant (e.g. -3/2 in Keplerian discs).
In non-axisymmetric discs the relation between $T^{B}_{r\phi}$ and $P_0$ is more complex.}.

The derived values of $\alpha$ vary significantly between the studies.
For instance, \citet{Hartmann98} derived $\alpha\approx0.01$ using observed disc sizes 
and a simple model for the evolution of an axisymmetric viscous disc, in which viscosity 
is a power-law function of radius.
\citet{Fleming} have employed a stratified model 
of accretion discs in a shearing box approximation, in which the upper layers are MRI-active 
while the central regions 
are quiescent. They found that $T^{B}_{xy}/P_0$ has a mean value of about 
$10^{-3}$ in the MRI-active upper layers but drops to a negligible value in the midplane.
On the other hand, $T^{Re}_{xy}/P_0$ in the vertical direction is roughly constant 
at a few $\times 10^{-4}$. This implies mean values of $\alpha$ in the range $10^{-4}-10^{-3}$.
Low mean values of $\alpha$ were also reported by Brandenburg et al. ($\alpha=0.007$) and 
Stone et al. ($\alpha\la0.01$). On the other hand, global numerical simulations tend to yield 
larger values for $\alpha$. For instance, \citet{Armitage} found mean values between 0.05 and 0.1.

Large variations in $\alpha$, both in space and time, imply that the development of the magneto-rotational
instability is strongly dependent on the local conditions in the disc. Nevertheless, we can still 
learn about circumstellar discs from a simple model used in our paper, if we assume that the 
constant $\alpha$ represents a mean value, time-averaged over many orbital periods of the disc.
Radial variations in $\alpha$ may (and should) be present in the disc but it requires a more 
thorough consideration of the disc physics (i.e. the ionization balance) and is left for a follow-up
paper.

Our numerical simulations unambiguously demonstrate that circumstellar discs cannot sustain
turbulent viscosity with a spatially and temporally averaged $\alpha \ga 0.1$. 
Such discs would have vanished during just one million year of evolution. The ubiquitous 
presence of older discs makes such large values of $\alpha$ unlikely. On the other hand,
low values of $\alpha$ of order $10^{-4}$ make little effect on the secular disc evolution,
which in this case is completely governed by gravitational torques rather than viscous ones.
In the case of $\alpha=10^{-3}$, viscosity does have some effect on the disc radial structure 
but the magnitude of these changes are modest -- the surface density profile becomes 
somewhat shallower in the inner disc and at the disc's outer boundary, 
and the disc size increases by a factor of 2 as
compared to that of the non-viscous one. 
The $\alpha=0.01$ disc sees considerable 
changes in its radial structure in the late evolution, and the gravitational
stabilization of such discs presents difficulty to account for
non-axisymmetric structure and poses problems for the theoretical idea of
giant planet formation via direct gravitational instability.
Moreover, the gas surface density in the entire $\alpha=0.01$ disc becomes 
lower than that of the MMSN after 1.0~Myr. This also poses problems for
planet formation models, which often require discs with gas surface densities 
a few times greater than that of the MMSN \citep[e.g.][]{Ida04}.
We conclude that the mean value of $\alpha$ (averaged over many orbital periods)
should lie in the range $10^{-3}-10^{-2}$, although large
transient variations around these values can still be present in real discs.

\subsection{Effective viscosity due to gravitational torques}

\begin{figure}
  \resizebox{\hsize}{!}{\includegraphics{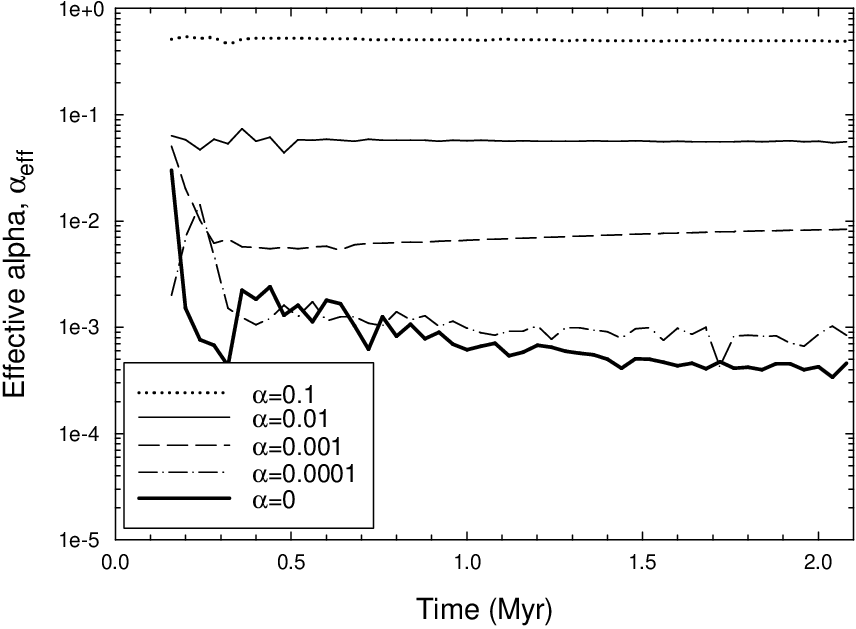}}
      \caption{Temporal evolution of the effective viscosity
parameter $\aleff$ that accounts for both viscous and gravitational
torques, for each of our models.}
         \label{fig11}
\end{figure}

An interesting way to account for the {\it combined} effect of 
gravitational and viscous torques is to calculate an
'effective alpha' $\aleff$ near the inner boundary of our 
simulation\footnote{For various methods to define $\aleff$ see e.g. \citet{Lodato07}}.
If we apply the steady-state mass accretion rate formula for 
thin viscous discs and also apply the $\alpha-$prescription, then
\begin{equation}
\dot{M} = 3 \pi \nu \Sigma = 3 \pi \aleff \tilde{c}_{s} Z \Sigma.
\end{equation}
We calculate $\dot{M}$ in each simulation at the boundary of the
sink cell ($r=5$ AU), and $\tilde{c}_{s}, Z,$ and $\Sigma$ at some distance
($r \approx 9$ AU) since $\Sigma$ decreases somewhat near the sink cell.
These numbers are used to generate $\aleff$, which is plotted versus time
in Fig. \ref{fig11} for each of our models. The values of $\aleff$ may
be scaled down by a factor $\approx 2$ if use also the values
of $\dot{M}$ at 9 AU instead of 5 AU. 
Clearly, gravitational torques in the absence of turbulent viscosity 
accounts for an 
$\aleff$ in the range $10^{-4}-10^{-3}$ during the late evolution.
This is why the addition
of a viscous $\alpha$ of at least $10^{-3}$ is required
to see significant changes in the disc evolution. 
  
It is also interesting to compare our results with previous numerical simulations of 
the secular evolution of viscous circumstellar discs. For instance, \citet{Lin}  
have considered the formation and evolution of a circumstellar disc formed
during the collapse of a rotating cloud core with initial mass $1.0~M_\odot$.  
They use the usual diffusion equation describing the evolution of the surface density
in a viscous {\it axisymmetric} accretion disc \citep{LBP}
\begin{equation}
{\partial \Sigma \over \partial t} = - {1\over r} {\partial \over \partial r} 
\left[ {1\over (r^2\Omega)^\prime}  {\partial \over \partial r} \left(  \nu \Sigma \,r^{3} \Omega^\prime 
\right) \right],
\end{equation}
complemented
by some form of the energy equation describing the internal energy balance in the disc
due to viscous heating, energy input from the accretion process, and disc radiative 
cooling. Our model, though neglecting detailed thermodynamics, accounts for a possible
disc asymmetry by directly solving the corresponding fluid dynamics equations 
for a thin disc. The effective kinematic viscosity in Lin \& Pringle's model comes 
from turbulent viscosity and 
gravitational instability, the former is parameterized using a usual Shakura \& Sunyaev 
$\alpha$-prescription (eq.~\ref{alpha}), while the latter is taken into 
account following \citet{Lin87}:
\begin{equation}
\nu_{\rm g} = \left\{ \begin{array}{ll} 
   {2\mu \over 3} \left( {Q^2_c  \over Q^2} -1 \right) 
   \left( {c^2_s \over \Omega} \right) & \,\, \mbox{if $Q \le Q_c$} \\ 
   0 & \,\, \mbox{otherwise}.  \end{array} 
   \right. 
   \label{alpha2}  
 \end{equation}
 We need to parameterize only the viscous torques,
the effect of gravitational torques is taken into account self-consistently.

Both approaches yield circumstellar discs that share some common characteristics.
For instance, Lin \& Pringle's model A1 ($\alpha,\mu=0.01$) produces a rather cold 
disc ($T\sim 10$~K) that features a near-flat surface density profile 
near the inner boundary and scales roughly as $r^{-1}$ at $100-1000$~AU.
Their disc has a sharp outer edge upon formation but it spreads out in the 
course of evolution.
These features are also seen in our modeling, though our disc in the corresponding 
model~3 has a smaller size ($\sim 300$~AU) and a lower surface density. This brings about  
the most striking difference found between the two approaches -- the derived disc masses. \citet{Lin} have 
reported disc masses that are an order of magnitude larger than the corresponding 
stellar masses in the early evolution. Although this large difference reduces with time,
the disc mass is still comparable to that of the star in the late evolution ($t\ga 1.0$~Myr), 
irrespective of the value of $\alpha$. Such massive 
discs are not observed. Our models predict maximum disc-to-star mass ratios of $\xi=14\%$ (model~1),
and this value quickly reduces with time for the viscosity-dominated discs (e.g. model~3).
Although our obtained values of $\xi$ for non-viscous models still seem to be greater than those usually inferred
for T Tauri stars, $0.5\%-1\%$ \citep[e.g.][]{Andrews}, 
they are not unfeasible given that the measured disc masses may be underestimated by 
conventional methods by as much as an order of magnitude
\citep[e.g.][]{Hartmann}. Figure~\ref{fig5} indicates that gravitational torques are 
a dominant mechanism
of mass and angular momentum transport in the disc in its early evolutionary phase.

The fact that \citet{Lin} obtain overmassive discs in their numerical simulations
suggests that either the parameterization given by equation~(\ref{alpha2}) is inadequate, 
certainly in the early disc evolution, or the ratio of rotational-to-gravitational energies $\beta$
in their model is too large, resulting in most of the initial cloud mass landing on to the disc
and through the disc on to the star rather than directly on to the star. 
In other words, the phase of near constant accretion
of matter from the envelope directly on to the star (see Fig.~\ref{fig6}) is very short
in the Lin \& Pringle model and the disc is not capable of processing the infalling mass 
to the star fast enough to keep its mass low. Indeed, Lin \& Pringle have adopted $\beta$ in the range
$0.25-0.64$, which is much larger than the values recently inferred for molecular cloud cores by 
\citet{Caselli}, $\beta=10^{-4}-0.07$. In our model, $\beta$ is set to $1.4\times 10^{-3}$.

\section{Summary}
\label{summary}
Using numerical hydrodynamics simulations we have studied the long-term evolution (2.0 Myr) 
of self-consistently formed, self-gravitating circumstellar discs that are subject to 
turbulent viscosity. We seek to determine the effect of 
viscosity on the radial structure and accretion properties of self-gravitating discs 
around low-mass ($\sim 0.7~M_\odot$) protostars.
We make no specific assumptions about the source of turbulence
in circumstellar discs and parameterize the magnitude of turbulent viscosity 
using the usual $\alpha$-prescription \citep{SS}. Four models with a spatially
and temporally uniform values of $\alpha=10^{-4}$, $10^{-3}$, $10^{-2}$, and $10^{-1}$
were considered and compared with the standard model characterized by $\alpha=0$. 
We find the following.
\begin{enumerate}
\item Low values of $\alpha$ of order $10^{-4}$ make little effect on the
secular evolution of a self-gravitating disc, the radial structure and accretion properties 
of which in this case are completely determined by gravitational torques rather than by viscous ones.
\item At values $\alpha$ of order $10^{-2}$, the discs see 
considerable changes in their radial structure,
with a surface density that is axusymmetric and has values that are already below 
that of the MMSN by 1 Myr. This is problematical for planet formation.
\item High values of $\alpha$ of order $10^{-1}$ make a catastrophic effect on 
the disc secular evolution. Most of the disc mass is quickly accreted on to the protostar and the rest
is dispersed to the external environment. The disc mean surface density drops below 
$1.0$~g~cm$^{-2}$ during just $1-2$~Myr.
\item  The net viscous torque in the disc is {\it positive} in the early evolution ($\la 0.1$~Myr)
and negative afterwards. On the other hand, the total (viscous plus gravitational) net torque is 
always negative, which is related to the removal of angular momentum from the disc
by gas that is accreted in to the central region.
\item Use of a stiffer barotropic equation of state ($\gamma=1.67$ instead 
of $\gamma=1.4$) and associated rise in disc temperature
can substantially affect the instantaneous accretion rates (particularly in the early evolution) 
but have little effect on the
disc accretion properties averaged over many disc orbital periods ($\sim 10^4$~yr). This is because
a decrease in the intensity of mass accretion bursts in the hotter disc is compensated by
an increase in the relative strength of lower order spiral modes ($m\le 2$), which are the most efficient
agents for radial mass transport in the disc. 
\item Viscous-dominated models have difficulty to account for some physical properties of 
circumstellar discs. For instance, they become virtually axisymmetric and 
gravitationally stable after just 1.0~Myr of evolution.
Moreover, the lack of a quiescent phase 
of low-rate mass accretion ($\dot{M}\sim 10^{-8}~M_\odot$~yr$^{-1}$) in the early  
evolution of viscous discs will make it difficult to account for Very Low Luminosity Objects 
(VeLLOs), which are presumably young objects that feature some combination
of a sub-solar mass and low accretion rate \citep{Young,Andre,Stecklum}. On the other hand,
non-viscous self-gravitating models can naturally account for both the apparent 
disc non-axisymmetry and a phase of very low luminosity.

We emphasize that the
$\alpha$-parameter in our models does not include a (possible) contribution 
from Reynolds stresses due to
self-gravity. This fact should be taken
into account when comparing our predicted values of $\alpha$ with those 
derived in other studies, which may include a contribution
from the gravitational Reynolds stresses \citep[e.g.][]{Gammie01}.
We have also not considered a possible contribution of Reynolds stresses
in the transport of mass and angular momentum.
According to \citet{Lodato04}, this contribution may be comparable to that 
of the gravitational torques, 
which means that the importance of self-gravity may be underestimated 
by a factor of 2.

\end{enumerate}

\section*{Acknowledgments} 
We are thankful to the referee, Giuseppe Lodato, for the insightful comments  that helped improve
the manuscript.
E.I.V. gratefully acknowledges present support from an ACEnet Fellowship. 
SB was supported by a grant from the Natural Sciences and Engineering
Research Council of Canada.
Numerical simulations were done 
   on the Atlantic Computational  Excellence Network (ACEnet) and Shared Hierarchical
Academic Research Computing Network (SHARCNET).

\appendix
\section{disc scale height}
We derive the disc vertical scale height $Z$ at each computational cell 
via the equation of local vertical pressure balance 
\begin{equation}
\rho \, \tilde{c}_s^2 = 2\int_0^Z \rho \left( g_{z,\rm gas}+g_{z,\rm st} \right) dz,
\label{eq1}
\end{equation}
where $\rho$ is the gas volume density, $g_{z,\rm gas}$ and $g_{z,\rm st}$ are the {\rm vertical} 
gravitational accelerations due to self-gravity of a gas layer and gravitational pull of 
a central star, respectively. 
Assuming that $\rho$ is a slowly varying function of vertical distance $z$ between $z=0$ (midplane)
and $z=Z$ (i.e. $\Sigma=2\, Z \,\rho$) and using the Gauss theorem, one can show that
\begin{eqnarray}
\int_0^Z \rho \, g_{z,\rm gas} \,  dz &=& {\pi \over 4} G \Sigma^2\:, \label{eq2} \\
\int_0^Z \rho \, g_{z,\rm st} \, dz &=& {G M_\ast \rho \over r} 
\left\{  1-\left[ 1+ \left({\Sigma\over 2 \rho r} \right) \right]^{-1/2} \right\},
\label{eq3}
\end{eqnarray}
where $r$ is the radial distance and $M_\ast$ is the mass of the central star.
Substituting equations~(\ref{eq2}) and (\ref{eq3}) back into equation~(\ref{eq1}) we obtain
\begin{equation}
\rho \, \tilde{c}_s^2 = {\pi \over 2} G \Sigma^2 + {2 G M_\ast \rho \over r} 
\left\{  1-\left[ 1+ \left({\Sigma\over 2 \rho r} \right) \right]^{-1/2} \right\}.
\label{height1}
\end{equation}
This can be solved for $\rho$ given the model's known 
$\tilde{c}_s^2$, $\Sigma$, and $M_\ast$, using Newton-Raphson iteration.
The vertical scale height is finally derived as $Z=\Sigma/(2\rho)$.

\section{Divergence of the viscous stress tensor}
The components of $\nabla \cdot {\bl \Pi}$ in polar coordinates ($r,\phi$) are
\begin{eqnarray}
\left( \nabla \cdot {\bl \Pi} \right)_r &=& {1\over r} {\partial \over \partial r} r \Pi_{rr} +
{1 \over r}  {\partial \over \partial \phi} \Pi_{r\phi} - {\Pi_{\phi\phi} \over r}, \\
\left( \nabla \cdot {\bl \Pi} \right)_\phi &=& {\partial \over \partial r} \Pi_{r \phi}
+ {1\over r} {\partial \over \partial \phi} \Pi_{\phi\phi} + 2 {\Pi_{r\phi} \over r},
\label{stress2}
\end{eqnarray}
where we have neglected the contribution from off-diagonal components $\Pi_{rz}$ and $\Pi_{\phi z}$.

\end{document}